\documentclass[aip,jcp,reprint]{revtex4-1}

\usepackage{dcolumn}       
\usepackage{bm}            
\usepackage{amsmath}       
\usepackage{graphicx}      
\usepackage{natbib}		   
\usepackage{hyperref}      
\usepackage{amssymb}       
\usepackage{latexsym}      
\usepackage{array}         
\usepackage{epstopdf}     

\newcommand{\bs}[1] {  \boldsymbol{#1}           }
\newcommand{\ff}[1] {  \mbox{\footnotesize{#1}}  }
\newcommand{\Ang}   {  \mbox{\normalfont\AA}     }

\begin{document}
\title{The origin of the Debye relaxation in liquid water and fitting the high frequency excess response }

\author{Daniel C.\ Elton}
\affiliation{Department of Physics and Astronomy, Stony Brook University, Stony Brook, New York 11794-3800, USA}
\affiliation{Institute for Advanced Computational Sciences, Stony Brook University, Stony Brook, New York 11794-3800, USA}

\date{\today}
\begin{abstract}
We critically review the literature on the Debye absorption peak of liquid water and the excess response found on the high frequency side of the Debye peak. We find a lack of agreement on the microscopic phenomena underlying both of these features. To better understand the molecular origin of Debye peak we ran large scale molecular dynamics simulations and performed several different distance-dependent decompositions of the low frequency dielectric spectra, finding that it involves processes that take place on scales of 1.5-2.0 nm. We also calculated the $k$-dependence of the Debye relaxation, finding it to be highly dispersive. These findings are inconsistent with models that relate Debye relaxation to local processes such as the rotation/translation of molecules after H-bond breaking. We introduce the \textit{spectrumfitter} Python package for fitting dielectric spectra and analyze different ways of fitting the high frequency excess, such as including one or two additional Debye peaks. We propose using the generalized Lydanne-Sachs-Teller (gLST) equation as a way of testing the physicality of model dielectric functions. Our attempts at fitting the experimental spectrum using the gLST relation as a constraint indicate that the traditional way of fitting the excess response with secondary and tertiary Debye relaxations is problematic. All of our work is consistent with the recent theory of Popov et al. (2016) that Debye relaxation is due to the migration of Bjerrum-like defects in the hydrogen bond network. Under this theory, the mechanism of Debye relaxation in liquid water is similar to the mechanism in ice, but the heterogeneity and power-law dynamics of the H-bond network in water results in excess response on the high frequency side of the peak. 
\end{abstract}

\maketitle

\section{Introduction}

The Debye relaxation peak dominates the dielectric absorption spectrum of water. The peak is centered at $\approx$ 20 GHz (0.66 cm$^{-1}$) and spans six decades of frequency. The large oscillator strength of the Debye peak (73 at 25 C$^{\circ}$) can be viewed as the main contributor to water's anomalously high static dielectric constant through the $f$-sum rule. The Debye absorption peak is of immense practical importance as it is used in microwave ovens and in satellite-based microwave radar sensing of water and ice.\cite{Rosenkranz2015:1387,Wang1980:288,Kneifel2014}  

Water is an anomalous, complex liquid that serves as the arena for all life on planet Earth. As such, there is a long legacy of research on water which continues to be built upon today. Current topics of interest among researchers include the possible liquid-liquid phase transition in water,\cite{Nilsson2016:8988,Limmer2011:134503} understanding water's behaviour under extreme thermodynamic conditions and confinement, and understanding the important role of nuclear quantum effects in water.\cite{Ceriotti2016review,PhysRevB.92.134105} In this paper we focus on the dielectric relaxation of water, about which an enormous literature already exists, including several recent detailed theoretical studies.\cite{Popov2016:13941,Ishai2015:15428,Hansen2016:237601,Arbe2016:185501,Lunkenheimer:1612.01457} In our review of this literature, we found a lack of agreement on the molecular origins of the Debye relaxation and further disagreement on how to fit excess response on the high frequency side of the Debye relaxation, which historically had been fit with a secondary Debye relaxation. Some authors relate Debye relaxation to particular translational and/or rotational motions of a single molecules after breaking one or more hydrogen bonds.\cite{Ronne1997:5319,Vij2004125,Agmon1996:1072} Other authors propose that Debye relaxation is due to the movement of ``free'' molecules which only have one or two hydrogen bonds.\cite{Buchner1999:57,Nabokov:1473,stanley:3404,Hasted1985:622,Zasetsky2011:025903,vonHippel1988} In contrast to these theories, other authors describe Debye relaxation as a collective relaxation of a cluster or large collection of molecules,\cite{Arkhipov:127,Arkhipov2003} which is more in line with dielectric theories that establish how the collective relaxation time increases due to dipole-dipole correlation. Finally, recently a few authors have also proposed the hopping of defects allow the rearrangement of the hydrogen bond network and Debye relaxation,\cite{Popov2016:13941,Ishai2015:15428,Artemov2014:158} in analogy to the process underlying Debye relaxation in ice.\cite{hobbs2010ice} As a macroscopic measurement, dielectric relaxation spectroscopy alone does not provide enough information to distinguish between the various mechanisms that have been proposed. In this work, we analyze large scale molecular dynamics simulations to gain insight into the collective nature of Debye relaxation.

It is worth dwelling on the fact that the Debye peak in water is nearly perfectly Debye. Most other dipolar liquids and virtually all polymeric liquids exhibit more complex relaxation.\cite{Kindt1996:10373} In such liquids, a phenomenological equation called the Havriliak-Negami (HN) formula is often used:
\begin{equation}\label{HNeqn}
   \frac{\varepsilon (\omega) - \varepsilon_\infty}{\varepsilon(0) - \varepsilon_\infty} = \frac{A_{\ff{HN}}}{[1+(i\omega\tau_{\ff{HN}})^\alpha]^\beta}
\end{equation}
The case $\alpha \neq 1, \beta = 1$ is known as Cole-Cole relaxation, and corresponds to a symmetric distribution of exponential relaxations. The applicability of the HN equation to water has been tested several times before. For instance Kaatze (1993) found $\alpha = .989(2)$ and $\beta = .959(4)$.\cite{Kaatze1993:95} Vij et al.\ (2004) report $\alpha = 1$, $\beta = 1$.\cite{Vij2004125} Mason's careful fit of the Cole-Cole equation yields $\alpha = 0.988 \pm .008$.\cite{Mason1974217}

The fact that the Debye relaxation peak is so well described by a single relaxation process clashes with our understanding of the structure and dynamics of the hydrogen bond network. Saito and Ohme \cite{Saito1994:6063} used molecular dynamics simulation to find that the relaxation of the polarization vector of a large water cluster exhibits a long-time tail, which can be fit with either a stretched exponential (also called a Kohlraush function, $P(t) \propto \exp [ -(t/\tau)^\beta ]$) or a $1/f^\alpha$ power law, both of which correspond to a broad distribution of relaxation times.\cite{babaSaito1997:3329} Similarly, molecular dynamics studies show that both the relaxation of the polarization vector of single molecules and hydrogen bond auto-correlation functions are well described by a stretched exponential.\cite{Luzar1996:55,Kumar:041505} Ohmine and Tanaka, in a detailed review, present evidence from molecular dynamics simulations that hydrogen-bond network rearrangement dynamics are complex and highly collective in nature.\cite{Ohmine1993:2545} They find evidence that the hydrogen bond network ``contains many relaxation processes, with many time scales''. This is reflected in the fact that the low frequency Raman spectrum of water between $1 - 20$ cm$^{-1}$ is very diffuse and can be fit with a power law.\cite{Ohmine1993:2545} Reconciling the complex heterogeneous dynamics of the H-bond network with the pure exponential character of the Deybe relaxation has been recognized as an important unresolved issue.\cite{Shiratani1996:7671,Yamaguchi2003:1211,Ohmine1993:2545}
A possible clue to solving this issue comes from the fact that if Coulomb interactions are smoothly truncated at 9 $\Ang$, dielectric relaxation decreases from $\approx 9$ ps to only 1 ps and assumes a $1/f^\alpha$ character.\cite{Ohmine1993:2545} The structure of water in such situations may be highly non-physical as well.\cite{Yonetani:49,Spoel:1} These results suggest that long range dipole-dipole interactions and/or the long range structure of the H-bond network are necessary to recover the exponential character of Debye relaxation. Further evidence comes from studies with salt, which show that Debye relaxation does not change very much with increasing salt concentration.\cite{Vinh2015:164502,Turton}


\section{Critical analysis of previous ideas about Debye relaxation}
\subsection{Models based off Debye-Stokes theory}
\label{sec:DebyeTheory}
We first consider Debye's original theory from 1929.\cite{D29} Debye considers a thermal ensemble of {\it non-interacting} molecules in an applied electric field, and considers what happens when the field is turned off. His starting assumption, which is now known to be incorrect, is that each individual water molecule undergoes Brownian rotational motion. Molecular dynamics simulations show that molecular relaxation in water actually occurs in highly discontinuous ``jumps" due to the breaking of H-bonds, rather than small angle Brownian diffusion.\cite{Luzar1996:55,Laage2008:14230,Laage2006:832,Ludwig2007,GEIGER2003131} Debye then takes the diffusion equation in spherical coordinates, linearizes it, and finds that the average moment decays exponentially with time, leading directly to the Debye equation for dielectric relaxation:
\begin{equation}
    \varepsilon (\omega)= \frac{\varepsilon(0) - \varepsilon_\infty}{1+i\omega\tau_D} +  \varepsilon_\infty
\end{equation}
Under this model, the Debye relaxation time $\tau_D$ is related to the rotational friction constant $\zeta$ via $\tau_D = \zeta/2k_B T$. Stokes showed that for a sphere of radius $a$ rotating in a medium with shear viscosity $\eta$, the rotational friction is given by $\zeta = 8\pi\eta R^3$. This leads to the Debye-Stokes model for $\tau_D (T)$:
\begin{equation}\label{DebyeStokes}
	\tau_{\ff{DS}}(T) = \frac{4\pi\eta(T) R^3}{k_B T}
\end{equation}
This equation fits the experimental data for $\eta(T)$ and $\tau_D(T)$ remarkably well, with a value of $R = 1.44 \Ang$ at 0 C$^\circ$ which is about the right radius for a single water molecule.\cite{Ronne1997:5319,Agmon1996:1072} Many authors have noted this agreement and concluded that Debye's model is essentially correct, and that Debye relaxation is due to the rotational relaxation of single molecules. 
Furthermore, Stoke's model leads to the Stokes-Einstein relation, which says that the translational diffusion constant is given by:
\begin{equation}\label{DiffDebyeRelation}
	\frac{1}{D(T)} = \frac{6\pi\eta(T) R}{k_B T} \propto \tau_{\ff{DS}}(T)
\end{equation} 
The Stokes-Einstein relation is borne out experimentally, as is the proportionality between $\tau_D(T)$ and $1/D(T)$ -- both follow Arrhenius-like temperature dependencies with very similar rate coefficients.\cite{Nabokov:1473} Equation \ref{DiffDebyeRelation} only breaks down when water comes supercooled.\cite{bertolini:3285,Agmon1996:1072,Arkhipov:127} Bertolini argues that the Arrhenius temperature dependence must be due to a barrier hoping process.\cite{bertolini:3285} Agmon builds his theory for Debye relaxation in water on equations \ref{DebyeStokes} and \ref{DiffDebyeRelation}, suggesting that it is due to translational hopping. In particular, he proposes that Debye relaxation is due to a hopping process called ``tetrahedral displacement''.\cite{Agmon1996:1072} Tetrahedral displacement has a hopping distance of 3.3 $\Ang$, which is ``the separation between an occupied and unoccupied corners of a cube binding the pentawater tetrahedron''.\cite{Agmon1996:1072}  We will argue later that Agmon's model is either incorrect or incomplete, because it does not explain the collective nature Debye relaxation that we find.

Hansen et al.\ note that both the Debye-Stokes and Stokes-Einstein equations fail for ordinary liquids, yielding a value $R_H$ which is much smaller than the molecular $R$.\cite{Hansen2016:237601} They note that for most molecular liquids, empirically it is found that $R_H \approx R/2$.\cite{Hansen2016:237601} If one uses this empirical relation, one finds $\tau_{\ff{DS}} \approx \tau_{D} / 8 \approx \tau_2$. In other words, the agreement of $R_H \approx R$ when equation \ref{DebyeStokes} is applied to water's Debye relaxation should be taken as coincidental. This makes sense if one understands $\tau_D$ as a collective phenomena, and $\tau_2$ as being related to single molecule relaxation.

\subsection{Mean-field theories}
The Debye-Stokes model is based on the flawed assumption that interactions between molecules can be ignored - notably dipole-dipole forces and H-bonding interactions. The effects of dipole-dipole interaction can be approximately accounted for by using a mean-field theory. The simplest mean-field theory is obtained by taking Debye's molecular dipole undergoing Brownian diffusion and moving it into a homogeneous medium described by a frequency dependent dielectric function $\varepsilon(\omega)$. The dipole then feels an additional ``internal field'' due to the polarization response of the medium. After solving the system self-consistently, one again obtains the Debye equations for $\varepsilon(\omega)$, but now:\cite{Arkhipov:127}
\begin{equation}\label{DebyeExtension}
	\frac{\tau_D}{\tau_s} = \frac{\varepsilon(0) + 2}{\varepsilon_\infty +2} \approx 11 
\end{equation}
In other words, the relaxation time for the polarization of the entire system, as measured through $\varepsilon(\omega)$, is greater than the relaxation time for the single dipole $\tau_S$. Here $\varepsilon_\infty$ can be understood as the excess oscillator strength not described by Debye relaxation. We use the experimental values $\varepsilon(0) = 78.6$ and $\varepsilon_\infty = 5.4$ to obtain a ratio of $10.8$. The ratio experimentally varies considerably with experiment (table I) between 8.3 - 34, with an average of $\approx 13$. We know that the Debye model is wrong, though, because the same mean field theory framework gives a completely wrong estimate of $\varepsilon(0)$, and predicts that water exists in a ferroelectric phase at room temperature. 

The Onsager mean-field model for $\varepsilon(0)$,\cite{O23} consisting of a dipole in a cavity, is considered a significant improvement over Debye's model. When the dipole moment is in a cavity, a ``reaction field'' field appears in addition to the internal field. Cole extended the Onsager mean-field model to the time dependent case, yielding a complicated dielectric function which corresponds to a non-exponential relaxation. Glarum extended Onsager's model to the time-dependent case,\cite{Glarum:1371} recovering the Debye equations with:
\begin{equation}\label{Glarumrelaxation}
    \frac{\tau_D}{\tau_s} = \frac{3\varepsilon(0)}{2\varepsilon(0) + \varepsilon_\infty}  = 1.46
\end{equation}
A similar theory by Powles yields:\cite{powles:633}
\begin{equation}\label{Powlesrelaxation}
    \frac{\tau_D}{\tau_s} = \frac{3\varepsilon(0) G_K}{2\varepsilon(0) + \varepsilon_\infty} 
\end{equation} 
Here the Kirkwood factor $G_K$ includes the effects of local dipole-dipole correlation.
\begin{equation}
   \frac{\tau_D}{\tau_s} = \frac{2\varepsilon(0)+ \varepsilon_\infty}{\varepsilon(0) + 2\varepsilon_\infty} = 1.8
\end{equation}
A variety of more sophisticated mean field theories have been developed, the details of which we will not recount here.\cite{,BagchiBook2012,Fatuzzo:729,Nee1970:6353,Arkhipov:127,Bagchi1990:455,Arkhipov:127} A common outcome of these models is that dipole-dipole correlations lead to the macroscopic dipole relaxation time of being longer than the molecular one. The same principle holds for clusters relaxing in a dielectric environment. 

\subsection{Propagation of defects}
Finally we come to the recent idea that Debye relaxation is due to the propagation of defects, as is the case in ice. There are four main defects in ice -- two charged ($H_3O^+$ \& $OH^-$) and two uncharged (Bjerrum L \& D defects). The propagation of these defects are responsible for Debye relaxation in ice, with the defect with the lowest energy barrier determining the timescale of the relaxation.\cite{petrenko1999physics} 

Popov et al.\ show that in liquid water the activation energies of charged and uncharged defects should be roughly the same, so distinguishing which may underlie Debye relaxation requires some additional analysis. Artemov \& Volkov propose that Debye relaxation is entirely due to the conduction of the charged defects.\cite{Artemov2014:158} Their model makes the claim that nearly 10\% of water molecules are ionized, which differs by six orders of magnitude from the accepted value of $K_w = 10^{-7}$.\cite{Artemov2014:158,Volkov2014:46004} This idea is inconsistent with classical molecular dynamics simulations, which can satisfactorily reproduce the dielectric response of water, even though they do not contain charged defects.\cite{Elton2014:124504,Sega2015:1539} Furthermore, Popov et al.\ note that the dielectric relaxation does not depend on pH, as it would in Artemov \& Volkov's model.\cite{Popov2016:13941} Popov et al. propose Debye relaxation is due entirely to Bjerrum-like defects, which carry an effective charge. The bifuricated hydrogen defect (Bjerrium D-like) results in excess positive charge locally and the bifuricated Oxygen defect (Bjerrium L-like) results in excess negative charge locally. If defects follow ordinary diffusive behaviour ($\langle r^2(t) \rangle = 6 D_{\ff{defect}} t$), then it is easy to show that the Debye equation for the dielectric response results.\cite{Popov2016:13941} The existence of defects Bjerrium L \& D defects (biffuricated bonds) in liquid water is supported the analysis of x-ray scattering data by Sciortino et al.\cite{Sciortino1990:3452} It is also supported by the molecular dynamics simulations of Laage \& Hynes which found bifuricated bonds lead to jump relaxation.\cite{Laage2006:832} We find the theory of Popov et al.\ attractive as previously we showed that the librational and OH-stretching dynamics of water are very similar to that of ice and that both originate from propagating phonon-like modes which travel through the hydrogen bond network.\cite{Elton2016:10193,EltonThesis} 

\section{The secondary Debye process and high frequency excess}
\begin{table*}\label{DebyeExptData}
	\begin{tabular}{@{\vrule height 8pt depth 0pt width 0 pt} l l l l l l l c c  }    
$\tau_{D}$(ps)&$\tau_{2}$(ps)&$\tau_{3}(ps)$& $f_{1}$& $f_{2}$ & $f_{3}$&range (cm$^{-1}$)&  method & ref  \\
\hline
  8.3     &  1.0     &          &  72      &  1.69(3) &         & .03 - 3    & DRS & Barthel, 1990\cite{Barthel1990:369}  \\
  8.4     &  1.1     &          &  72      &  1.75    &         &.006 - 14   & DRS & Buchner, 1997\cite{Buchner1999:57}  \\
  8.4     &  0.91    &          &  72(1)   & 1.77(6)  &         & .075 -10   & DRS & Peacock, 2009\cite{Peakcock2009:205501}\\
  8.3     &  0.39    &          &  75      &  1.67(3) &         & .2 - 4     & DRS & Sato, 2008\cite{Sato2004:5007}  \\
  8.3     &  0.36    &          &  72      &  2.12    &         & 6 - 83     & ATR & M\"{o}ller,2009 \cite{Moller2009:A113} \\
          &  0.248(8)&          &  75(1)   & 1.67(3)  &         &            & ATR & Yada, 2008\cite{Yada2008:166}    \\
 7.0(3)   &  0.92(6) &          &  70(1)   & 2.0(3)   &         & 2 - 66     & TDS & Ronne, 1997\cite{Ronne1997:5319}\\
 8.3      &  0.42    &          &  73      &  2       &         & .001 - 3   & TDS & Fukasawa, 2005\cite{Fukasawa2005:197802} \\
 8.24(4)  &  0.18(14)&          &  73      &  1.9(5)  &         &  2-50      & fLS & Kindt, 1996\cite{Kindt1996:10373} \\
 8.8(6)   &  0.21(6) &          &  73      &  1.5(8)  &         &  3-55      & fLS & Venables, 1998\cite{Venables1998:4935} \\	
 7.8      &  0.2     &          &  73      &  1.6     &         & .16-33     & var & Liebe, 1991 \cite{Liebe1991:659} \\
  8.21    &  0.39(5) &          &  73      &  2.5(2)  &         & .1 - 33    & var & Benduci, 2007\cite{Beneduci2008:55}\\
  8.31	  &  1.0$^{*}$ &  0.10$^{\dagger}$&71.5   &  2.8     & 1.6     & 50-220     & dFTS& Vij, et al.\ 2004\cite{Vij2004125}\\  
  8.26(3) &  1.1(5)  &  0.14(4) &  73      &  2.2(2)  & 1.3(3)  & .1 - 33    & var & Benduci, 2007\cite{Beneduci2008:55}\\
  8.5	  &  0.93    &  0.08    &          &          &         & .03-800    & var & Ellison, 2007\cite{ellison:1}  \\  
  8.4(3)  &  1.05(15)&  0.18(5) &          &          &         & 0.02-37    & var & Vinh, et al.,\ 2015\cite{Vinh2015:164502}  \\
      \end{tabular}
      \caption{Reported two-Debye and three-Debye fits for experimental data taken at 298 K (25 C) over the last 30 years. DRS = microwave dielectric relaxation spectroscopy, ATR = THz attenuated total reflectance spectroscopy, TDS = THz time domain reflection spectroscopy, fLS - femotosecond laser spectroscopy, dFTS = dispersive Fourier Transform Spectroscopy
      \\$^*$HN model for $\tau_2$, $\alpha = 1, \beta = .77$ \quad $^{\
      \dagger}$ HN model for $\tau_3$, $\alpha = .9, \beta = .8$  }
\end{table*} 

We also seek to address an ongoing controversy on how to fit the high frequency side of the Debye relaxation, between 1 - 100 cm$^{-1}$, where there is unaccounted for excess response. Ishai et al. and Popov et al. propose that excess is also due to the dynamics of defects propagating through the hydrogen bond network which was discussed in the previous section.\cite{Ishai2015:15428,Popov2016:13941} Traditionally however, this excess response has been fit by introducing a second Debye mode, characterized by a time constant $\tau_2$. However, while the value of $\tau_D$ is very consistent among experiments, values for $\tau_2$ vary considerably in the experimental literature, as shown in table I. An additional problem, noted by Beneduci, is that the data on the temperature dependence of $\tau_2$ is contradictory - Barthel et al.\ find it increasing with temperature while Ronne, et al. find it decreasing.\cite{Beneduci2008:55} To fix these issues, recently it was proposed that a 3rd Debye relaxation ($\tau_3$) is also required to properly fit the high frequency excess.\cite{Beneduci2008:55,Vinh2015:164502}

As with the primary Debye mode, different authors have different hypotheses about the microscopic mechanisms that underlie the secondary and putative tertiary Debye modes. $\tau_2$ is usually associated with the rotational relaxation of single molecules and hydrogen bond breaking. Some attribute it to the relaxation of weakly bound molecules.\cite{Buchner1999:57} Molecular dynamics show the average hydrogen bond lifetime to be around 0.5 - 1.0 ps at 300 K,\cite{Ohmine1995:6767} with a very broad distribution.\cite{Sciortino1990:1686} Others propose that the excess response is either the $\alpha$ or (more oftenly) $\beta$ relaxation found in supercooled liquids.\cite{Lunkenheimer:1612.01457}

In light of the heterogeneous and non-exponential dynamics of the hydrogen bond network, the use of one or two additional exponentials to fit the excess response seems ad-hoc. The ad-hoc nature of this fit can also be seen by considering some of the infrared active H-bond vibrational modes that have been shown to exist in the region of $10 - 200$ cm$^{-1}$, as shown in table II. Heyden et al.\ have shown that all of these modes span a broad frequency range between $20 - 200$ cm$^{-1}$ due to the inhomogeneous nature of the H-bond network.\cite{Heyden2010:12068} Instantaneous normal mode analysis of water shows a broad spectrum of translational modes, extending from very 1-400 cm$^{-1}$ and peaked around 100 cm$^{-1}$.\cite{Cho1994:6672,Ohmine1995:6767,Kindt1996:10373} 

Thus it appears that a large number of Debye and resonance processes contribute to the excess response. This can be modeled by fitting a distribution of modes. Unfortunately, fitting a distribution of modes is a mathematically ill-posed problem - many distributions and combinations of modes may be consistent to the experimental data within the uncertainty of the data. The wide variation in experimental fits is at least partially explained by the fact that the approximation of one or two additional Debye relaxations to fit the excess is crude and that each experiment only measures a certain window of frequencies. Since the fit is approximate, it varies depending on the particular range of the experiment. Furthermore, some authors do not unbias their fitting, so the fit can be biased toward the side of the spectrum that is higher in magnitude. Dielectric relaxation spectroscopy (DRS) experiments can only probe the low frequency part (.0001 - 2 cm$^{-1}$), while THz or microwave time-domain reflection spectroscopy (TDS) probe the``middle" frequencies (1 - 10 cm$^{-1}$), as does a variation known as attenuated total reflectance spectroscopy (ATR).\cite{Moller2009:A113} Finally, Fourier transform infrared spectroscopy (FTIR) covers the region above 10 cm$^{-1}$.  

A side point is that an excess response on the high frequency side of Debye relaxation is a general feature found in many dipolar liquids, including non H-bonding liquids. It seems largely forgotten that this was first pointed out in 1955, when Poley noticed that an excess around 10 cm$^{-1}$ (0.3 THz) exists in many dipolar liquids.\cite{Poley1955:337} To explain this phenomena, which was called ``Poley absorption'' at the time, Hill and others proposed that it was due to inertial motion.\cite{Hill1971:2322,H69} Physically, inertial motion can be pictured as either fast ``rattling" of molecules within sharply defined potential energy wells or as nearly-free rotations over small angles.\cite{Bagchi1993:133,ferraro1978} In water, inertial absorption (if relevant) would overlap with hydrogen bond network vibrations and modes. Inertial relaxation results in approximately a Gaussian form for $\phi(t)$ near $t = 0$.\cite{Hill1971:2322} In our previous work\cite{Elton2016:10193} we did not find any evidence of such Gaussian relaxation except in $\phi(k,t)$ for $k > 3 \Ang^{-1}$. 

\begin{table}\label{Hbondmodes}
    \centering
    \begin{tabular}{c c c}
 approx freq. (cm$^{-1}$) & description         & ref\\ 
 \hline
    50-65   & H-bond bending (in plane)         &  \cite{Agmon1996:1072}      \\
        70  & H-bond torsion                    &  \cite{Agmon1996:1072}      \\
        150 & H-bond sym. stretch (``breathing")&  \cite{Agmon1996:1072}      \\
        180 & H-bond asym. stretch              &  \cite{Agmon1996:1072}      \\
    80-150  & assymetric umbrella mode          &  \cite{Heyden2010:12068}\\
    \end{tabular}
    \caption{Some of the H-bond network modes in liquid water. All of these modes are IR active and should appear in the dielectric response.}
\end{table}



\section{Fitting the temperature dependence}
\begin{figure}
 \includegraphics[width=8cm]{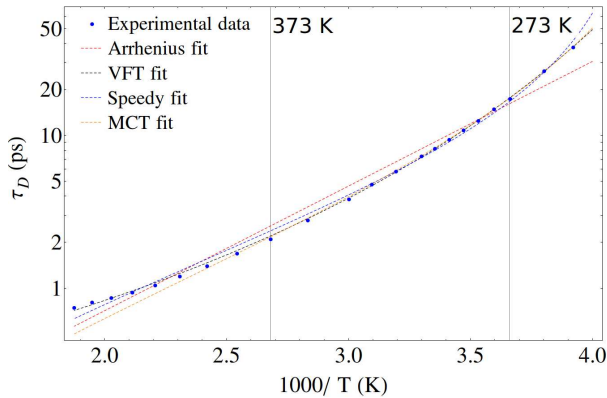}
  \caption{The temperature dependence of Debye relaxation. Data taken from Ellison (2007)\cite{ellison:1} and Nabokov (1988).\cite{Nabokov:1473} }
  \label{TauVsT}
\end{figure}
The temperature dependence of Debye relaxation is usually cited as being Arrhenius.\cite{Agmon1996:1072} According to transition state theory, Arrhenius temperature dependence implies an free energy barrier $\Delta H$, and in water one finds $\Delta H$ is roughly the hydrogen bond energy. On the basis of this, Buchner proposes that the Arrhenius temperature dependence of $\tau_D(T)$ is due to the production rate of ``free molecules'', which he defines as molecules having 1 or 0 H-bonds.\cite{Buchner1999:57} The percolation model of Stanley \& Tiexera indicates that $\tau_D(T)$ is due to reorientation of molecules having only one hydrogen bond.\cite{stanley:3404} Nabokov and Lubinov adapt a similar type of model, but argue that molecules with both one and two hydrogen bonds are mobile and thus contribute.\cite{Nabokov:1473} 

Nabokov et al.\ note that $\tau_D(T)$ deviates from Arrhenius behaviour at low temperature.\cite{Nabokov:1473} To our knowledge a test of different fit functions over the entire experimentally accessible range has not yet been published. We combined data from Ellison (2007)\cite{ellison:1} from between 273 - 373 K and the data collected by Nabokov (1988)\cite{Nabokov:1473} from between 255 - 533 K. The data taken above 373 K was measured along the gas-liquid co-existence curve. We fit the log normalized data with the following fit functions: 
\begin{equation}
    \begin{aligned}
        \tau(T) & = A_{\ff{a}} \exp\left( \frac{\Delta H}{kT} \right)        & \mbox{Arrhenius} \\
        \tau(T) & = \tau_{\infty} \exp \left( \frac{DT_{\ff{VFT}}}{T-T_{\ff{VFT}}} \right)     & \mbox{Vogel-Fulcher-Tammann} \\
        \tau(T) & = \frac{A_{\ff{s}}}{T}\left(\frac{T}{T_s} - 1\right)^\gamma& \mbox{Speedy's eqn.\cite{speedy:851} }\\
        \tau(T) &= A_{\ff{MCT}}*(T - T_{\ff{MCT}})^\gamma   & \mbox{mode-coupling theory\cite{Kumar:041505}} 
    \end{aligned}
\end{equation}

The results are shown in an Arrhenius plot in figure \ref{TauVsT}. Clear deviations from Arrhenius behaviour (which would be a straight line in this plot) are observed. The behaviour of $\tau_D(T)$ is very nearly Arrhenius between 273-313 K, but the only model which fits the data through the entire temperature range is the Vogel-Fulcher-Tammann relation, which is used to fit the ``$\alpha$-relaxation" in glasses. We found $\tau_{\infty} = 0.14$ ps, $D = 4.52$ and $T_{\ff{VFT}} = 141$ K. For the Arrhenius model we found $\Delta H = 0.16$ eV (3.7 kcal/mol), which is approximately the hydrogen bond energy. The mode-coupling theory power law was previously shown to fit $\tau_D(T)$\cite{Kumar:041505} in simulations of SPC/E water.\cite{PhysRevLett.82.2294} The Speedy equation was included because it is used to fit other response functions for water, especially in the supercooled region.\cite{speedy:851} 

VFT temperature dependence is a universal feature of both relaxor ferroelectrics and dipolar glasses,\cite{Pirc:020101,Bokov:4899} and indicates the presence of spatial heterogeneity.\cite{Vilgis:3667,Giovambattista2003:085506}  A phenomenological theory by Tagantsev shows that near $T_{\ff{VFT}}$ VFT temperature dependence is a consequence of a very wide distribution of relaxation times in the system.\cite{Tagantsev1994:1100} A very general theory for the VFT equation is the Adam-Gibbs model, which assumes the existence of cooperatively rearranging regions (CRRs), which are clusters that relax independently from each other, similar to the polar nanoregions concept we introduced in a previous work.\cite{Giovambattista2003:085506}

\section{Molecular dynamics simulations}
We performed a number of molecular dynamics simulations using the TIP4P/2005 water model\cite{abascal:234505} and the GROMACS 4.5.5 molecular dynamics package.\cite{abascal:234505} Our simulations used a Nos\'{e}-Hoover thermostat with $\tau = 0.5$ ps and a timestep of 2 fs. Long range Coulomb interactions were handled with the particle mesh Ewald method. All simulations were equilibrated for at least 100 ps before trajectory output. Our 10,000 molecule (box size $L = 6.68$ nm) simulation was 4 ns long and used a Coulomb cutoff of $r = 3.34$ nm. We analysed the simulation trajectories using our {\it epskw} Fortran code for molecular liquids. The code is open source and available online at \href{https://github.com/delton137/epskw}{www.github.com/delton137/epskw}. 
 
\section{$k$-dependence}
\begin{figure}
 \includegraphics[width=8.5cm]{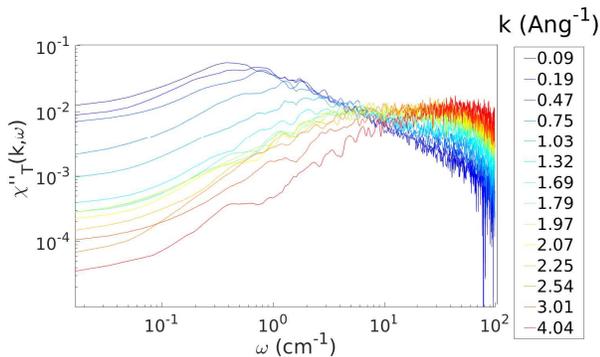}
  \caption{Imaginary part of the $k$-dependent transverse dielectric susceptibility for a box of 10,000 TIP4P/2005 molecules at 300 K showing the dispersion of the Debye peak.}
  \label{kdepDebye}
\end{figure}
\begin{figure}[h]
    \centering
        \includegraphics[width=8.5cm]{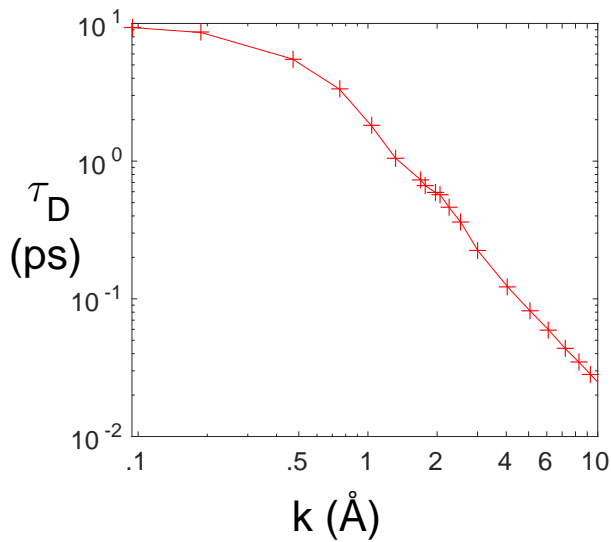}
  \caption{$\tau_D$ vs $k$ for a box of 10,000 TIP4P2005 water molecules.}
  \label{tauvsk} 
\end{figure}

Figure \ref{kdepDebye} shows the $k$ dependence of the Debye relaxation, calculated using the method we describe in our previous work.\cite{EltonThesis,Elton2016:10193} By fitting the underlying $k$ dependent correlation functions we produced a plot of $\tau_D(k)$ which is shown in figure \ref{tauvsk}. At larger spatial scales (smaller $k=2\pi/\lambda$) the relaxation is slower than at shorter scales (larger $k$). 

Recently, Arbe et al.\ have measured the dynamic structure factor of water, $S(Q,\omega)$ using incoherent and coherent neutron scattering in the GHz-THz frequency range and the intermediate $Q$ range (0.3 - 2.0 $\Ang^{-1}$).\cite{Arbe2016:185501} They observe a structural relaxation process at same frequency as the Debye peak in the dielectric spectra. Assuming the two processes correspond, our result (fig. \ref{tauvsk}) is qualitatively consistent with their measurement of $\tau_D(k)$ they derived from fitting a double exponential process to $F(Q,t)$, the intermediate incoherent scattering function for $H$ nuclei. Arbe et al.\ conclude that the $k$ dependence indicates the presence of translational diffusion, but is not consistent with an interpretation in terms of a single diffusive motion, as in Agmon's model. They label the Debye mode as the ``diffusive'' model, and the region of excess response is fit with a ``local'' Debye relaxation. Arbe et al.\ also found that the ``local'' Debye relaxation does not exhibit dispersion with $k$, (hence the name ``local'').\cite{Arbe2016:185501}

\subsection{Distance decomposition of the Debye relaxation}
\begin{figure}[h]
   \centering
        \includegraphics[width=8.5cm]{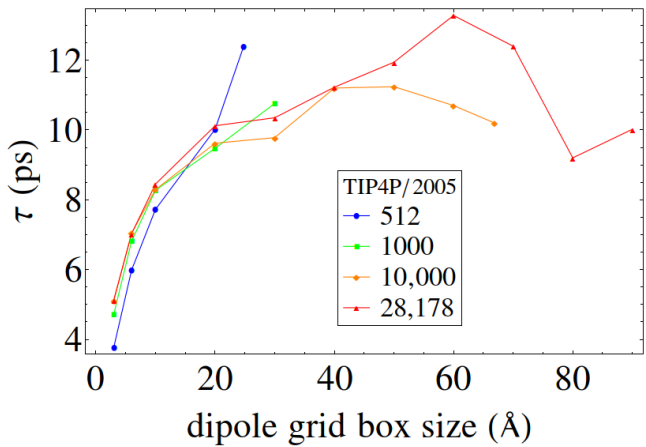}
  \caption{Debye relaxation calculated using the dipole grid method for simulations of TIP4P/2005 water with different simulation box sizes. Dipole grid box sizes of $L \approx 2$ nm are required for convergence.}
  \label{DipoleGridBoxes}
\end{figure}

There are several ways to perform distance-dependent decomposition of the Debye relaxation. The most intuitive way is to break the simulation cell into sub-boxes of different sizes and compute the total dipole moment for each sub-box. Dipole time correlation functions are then computed separately for each sub-box and averaged. The process is repeated for sub-boxes of different sizes (fig.\ \ref{DipoleGridBoxes}). We find that the dipole relaxation time does not converge to the bulk value until the box size is increased to $\approx 2.0$ nm.

We investigated two other methods of distance-decomposition. In the first method, which we call the ``dip-sphere" method, one starts with the dipole-dipole time-correlation function: 
\begin{equation}
    \phi(t) = \left\langle \sum_i \bs{\mu}_i(0) \cdot \sum_j \bs{\mu}_j(t) \right\rangle
\end{equation}
Then, one limits the molecules around each molecule $i$ to those in a sphere of radius $R$:
\begin{equation}
\phi^{\ff{ds}}(t,R) = \left\langle \sum_i \bs{\mu}_i(0) \cdot \sum_{j \in R_i} \bs{\mu}_j(t) \right\rangle
\end{equation}

The other method, which we call the ``sphere-sphere'' method, was introduced by Heyden, et al.\cite{Heyden2010:12068} The sphere-sphere method is so-called because one calculates the autocorrelation of the total dipole moment of a sphere of radius $R$ centered around a reference molecule, and then averages this over each molecule in the system:
\begin{equation}
\phi^{\ff{ss}}(t, R) = \sum_i \left\langle \bs{\mu}_i^s(0) \cdot \bs{\mu}_i^s(t) \right\rangle
\end{equation}
where:
\begin{equation}
	 \bs{\mu}_s(t) = \mathcal{N}_i(t)\sum_{j \in R_i} \bs{\mu}_j(t) 
\end{equation}

Heyden et al.\ recommend the normalization factor $\mathcal{N}_i(t) = (1 + N_{ij}^2)^{-1/2}$ to normalize for number of molecules in each sphere. This normalization factor is chosen so that in the bulk limit ($R \rightarrow \infty$) the original full response function is obtained. Heyden et al.\ also propose introducing a smoothing function to weight molecules within the sphere, which we neglect here for simplicity.

\begin{figure}[h]
    \centering
        \includegraphics[width=8.5cm]{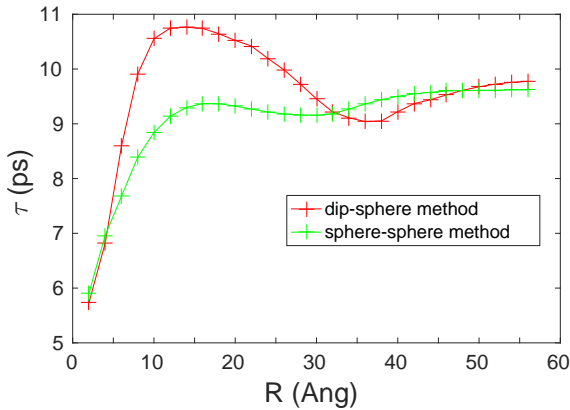}
  \caption{$\tau_D$ vs $R$ for a box of 10,000 TIP4P2005 water molecules using the ``dip-sphere'' and ``sphere-sphere'' methods.}
  \label{DipSphereDistDecomp} 
\end{figure}

The results of these two types of decomposition are shown in figure \ref{DipSphereDistDecomp}. The dielectric relaxation time increases up to about 1.5 nm. Interestingly, this is the maximum length scale we found feasible for clusters in liquid water -- beyond 1.5 nm water-water interactions behave as in a dielectric continuum.\cite{Elton2014:124504} A sphere of 1.5 nm contains approximately 470 water molecules. Our results indicate that complexes of 100s of molecules participate in Debye relaxation. The two models that are most consistent with this finding are the propagation of defects model and the model of relaxing clusters. 


\section{Fitting the high frequency excess}
As mentioned, there is uncertainty and confusion on how to fit and interpret the excess response on the high frequency side of the Debye relaxation. There are many pitfalls when fitting a dielectric spectra, especially when there there are many overlapping modes. When the underlying relaxation is described by several overlapping  modes or a distribution of relaxations and/or damped harmonic oscillators, the number of parameters becomes cumbersome or impossible for many fitting algorithms. More importantly, the fitting becomes fundamentally under-defined -- many fits become consistent with the data, within the noise of the data. Trying to fit a distribution of relaxation times $G(\tau)$ to a spectrum is an ill-defined problem, and attempts to do so can yield peculiar results.\cite{Zasetsky2011:025903} Still, this type of fitting can be done using Tichonov regularization, which enforces the distribution to be smooth and prevents overfitting.\cite{Schafer:2177,Dias:14212,Weese1992:99,Tuncer2006:074106,Jang2003:956}

Distinguishing the validity of various proposed dielectric functions is difficult. Sheppard and Grant found that ``.. data represented by a small departure from single relaxation time behaviour ($0.9 < \alpha < 1$ in eqn.\ \ref{HNeqn}) could be equally well interpreted as being due to two distinct kinds of relaxation process with relaxation times separated by a factor as high as three.''\cite{Sheppard1974:61} A similar point is made by Barker, who notes that a single Debye mode is nearly indistinguishable from two closely overlapping modes.\cite{Barker1975:4071}

Puzenko, et al.\ introduced the following dielectric function,\cite{Puzenko2005:6031}
\begin{equation}
    \begin{aligned}
    \varepsilon (\omega) &= \frac{f[1 + f(\omega)]}{(1+i\omega\tau_D)^\beta} + \varepsilon_\infty\\
    \mbox{where  }&\\
    f(\omega) &= \begin{cases} 0 \mbox{ if } \omega < \omega_c \\
                            A(\omega\tau)^q \mbox{ if } \omega > \omega_c \end{cases}        \end{aligned}
\end{equation}
This function was recently used by Ishasi et al.\ to fit the high frequency excess  in water.\cite{Ishai2015:15428} A cutoff frequency $\omega_c$ is required, since otherwise this expression violates the Kramers-Kronig relations. In our experiments with this fit function, we found this cutoff could be ignored. Putting the cutoff on at the center of the Debye peak results in an unphysical discontinuity in the fit function. As a power law, this fit corresponds to a flat distribution of relaxation times in the hydrogen bond network. The interpretation of $A$ and $q$ in this fit function are not clear. 

Under the defect-migration theory for Debye relaxation introduced by Popov, et al., the Debye peak assumes the following form:\cite{Popov2016:13941} 

\begin{equation}
       \varepsilon (\omega)  = \frac{f}{1 + \left[ (i\omega \tau_{\ff{defect}})^{-1} + (i\omega \tau_{\ff{osc}})^{-\delta} \right]^{-1}   } + \varepsilon_\infty
\end{equation}
This fit takes into account the high frequency wing, without the need for the addition of one or two extra Debye relaxation processes. Hidden in this dielectric function is a power law of the form $\omega^{-\delta}$, where $0 < \delta <1$.

A custom Python package called \textit{spectrumfitter} was developed, which is available at \href{https://github.com/delton137/spectrumfitter
}{github.com/delton137/spectrumfitter} or from the Python Package Index. We started with the assumption that dielectric function has $N$ Debye relaxation processes and $M$ damped harmonic oscillator processes, and thus has the following form: 
\begin{equation}\label{generalepsomega}
    \varepsilon(\omega) = \sum\limits_i^N \frac{f_i}{1+i\omega\tau_{Di}}  + \sum\limits_j^M \frac{f_j\omega_{Tj}^2}{\omega_{Tj}^2 - \omega^2 - i\omega\gamma_j}  + \varepsilon_\infty 
\end{equation}
Fitting is performed with the ``$f$-sum rule'' enforced as a constraint:
\begin{equation}\label{fsum}
   \varepsilon(0) - \varepsilon_\infty = \sum_i f_i
\end{equation}
where $f_i$ is the oscillator strength of the $i$th mode. We fit the experimental refractive index data compiled by Segelstein (1981), which comes from a compilation of all experimental sources of dielectric function and index data that were available at the time.\cite{Segelstein} While dated, the Segelstein dataset has the benefit that it covers the entire frequency range from $0.001 - 200,000$ cm$^{-1}$ The dielectric permittivity is obtained from the complex index of refraction ($n, k$) data using: 
\begin{equation}\label{indexrelation}
	\begin{aligned}
		\varepsilon'(\omega)  &= n^2(\omega) - k^2(\omega)\\
		\varepsilon''(\omega) &= 2n(\omega)k(\omega) 
	\end{aligned}
\end{equation}

We first did an interpolation of the Segelstein data on a logarithmic grid up to 18 cm$^{-1}$ and a linear grid from 18-4000 cm$^{-1}$, to prevent biasing the fitting towards the Debye mode. We performed the fitting in an unbiased manor by minimizing the sum of the relative errors squared: 
\begin{equation}
    \mbox{Cost} = \sum\limits_i \left(\frac{ \mbox{fit}_i - \mbox{data}_i}{\mbox{data}_i}\right)^2
\end{equation}

\begin{figure}[h]
    \includegraphics[width=7.2cm]{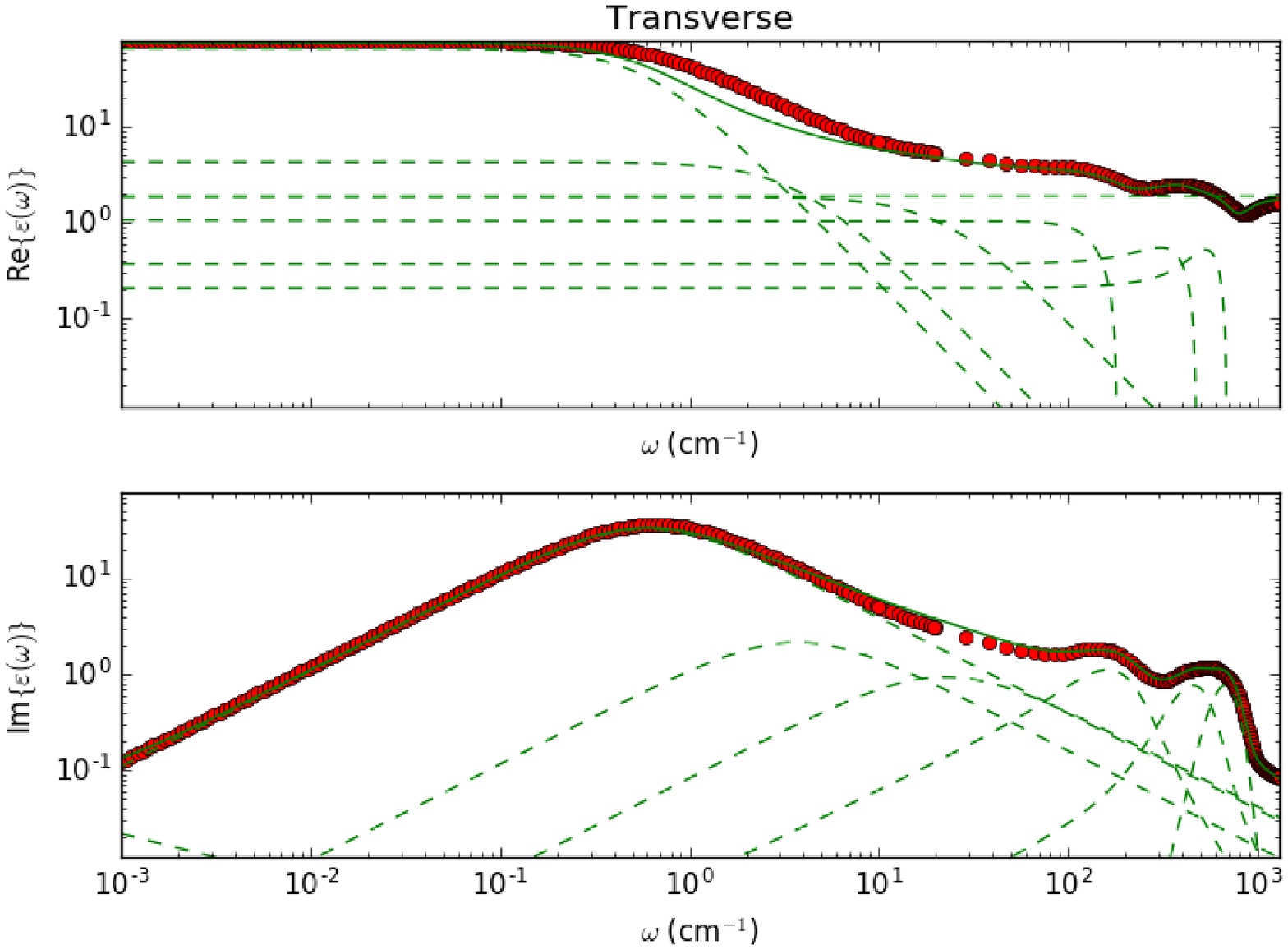}
    \includegraphics[width=7.8cm]{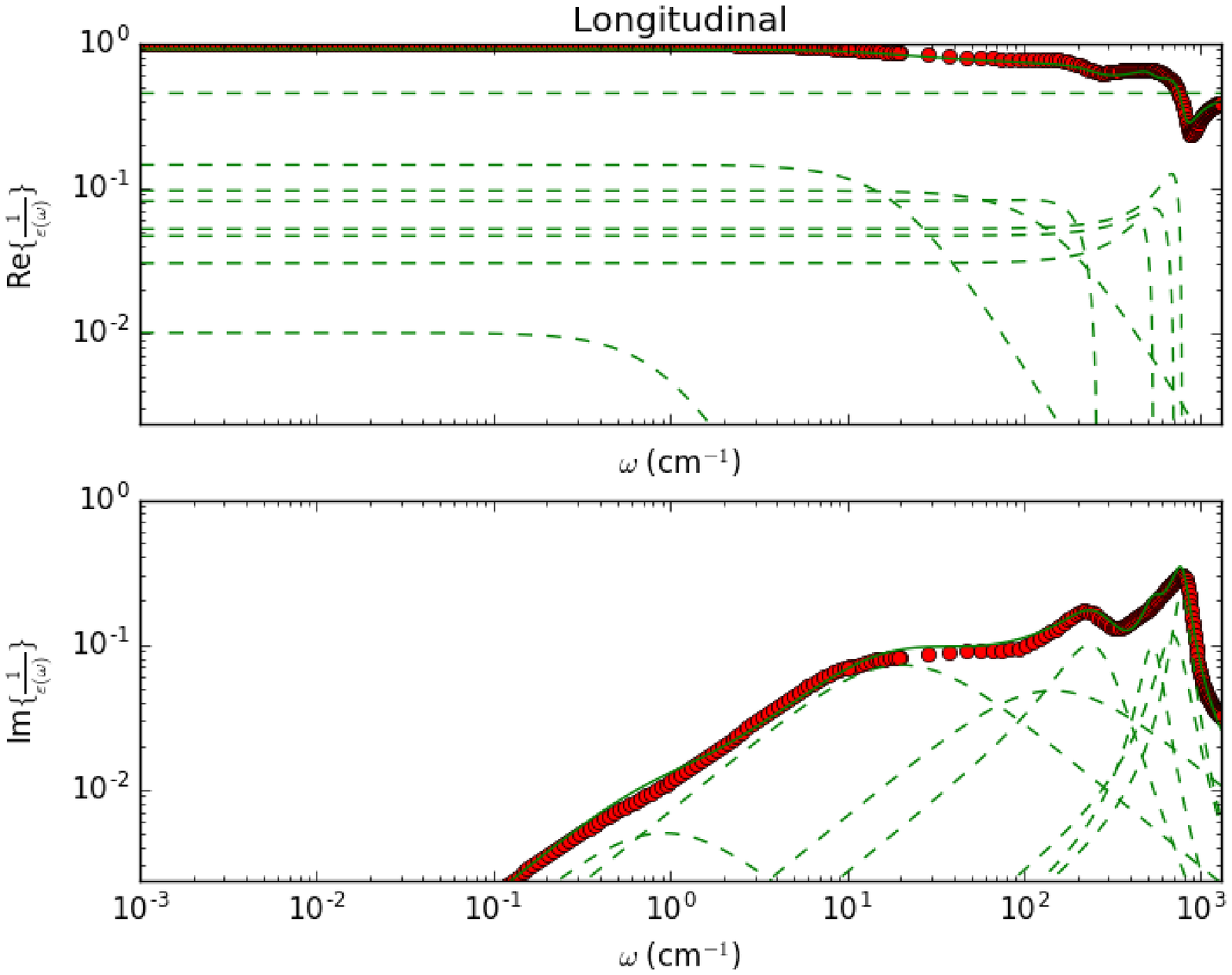}
     \begin{tabular}{c c c c c c}
name (Transverse) &  $f$ & $\omega_0$ & $\tau$ (ps) &  $\gamma$  &  $\sigma$ \\ 
\hline
               Debye & 65.0 &   0.60 &  8.88 &   & \\
           2nd Debye & 4.27 &   3.71 &  1.43 &   & \\
           3rd Debye & 1.86 &  22.5 &  0.24 &   & \\
   Brendel Hbond Str & 3.43 & 180 & 0.186 & 117  &  53.2 \\
          Brendel L1 & 1.28 & 474 & 0.070 & 107  & 120 \\
          Brendel L2 & 0.747 & 691 & 0.048 & 100  & 150 \\
          Brendel L3 & 0.010 & 689 & 0.048 & 227  &  46.8 \\
             eps inf & 1.88 & & & & \\
name (Longitudinal)            & $f$ & $\omega_0$ & $\tau$ (ps) &  $\gamma$ &  $\sigma$  \\ 
\hline
               Debye & 0.01 &   0.9 &  5.72 &   & \\
           2nd Debye & 0.14 &  20.3 &  0.26 &   & \\
           3rd Debye & 0.10 & 146 &  0.04 &   & \\
   Brendel Hbond Str & 0.31 & 259 & 0.129 & 221  &   2.85 \\
          Brendel L1 & 0.18 & 704 & 0.047 & 284  &  12.1 \\
          Brendel L2 & 0.20 & 782 & 0.043 & 183  &   7.20 \\
          Brendel L3 & 0.12 & 538 & 0.062 & 163  &  21.7 \\
           eps inf L & 0.45 & & & & \\
     \end{tabular}
    \caption{Example transverse (top) and longitudinal (bottom) fits and parameters. $\omega_0$, $\gamma$ and $\sigma$ are all reported in cm$^{-1}$. The fit contained 3 Debye relaxations, 1 Brendel peak for H-bond stretching, and 3 Brendel peaks for the librational region. The RMS error was 0.120. }
    \label{gLSTfit}
\end{figure}

\subsection{Using the gLST relation as a novel constraint}

Any dielectric function should obey the Kramers-Kronig relations, which is derived from the basic principle of causality (causes must proceed effects). The $f$-sum rule (eqn.\ \ref{fsum}) comes from taking $\omega = 0$ in the Kramers-Kronig relations. Barker shows how a generalized Lydanne-Sachs-Teller (LST) relation can also be derived from the Kramers-Kronig relations.\cite{Barker1975:4071}
For a single damped harmonic oscillator mode, the gLST relation is: 
\begin{equation} 
  \frac{(\omega^2_{L}+\gamma^2_{L})}{\omega^2_{T}} = \frac{\varepsilon(0)}{\varepsilon_\infty}
\end{equation}
The dampening factor appears in the numerator since the longitudinal frequency is complex $\bar{\omega}_L = \omega_L + i \gamma_L$. The generalized LST relation for $N$ Debye modes and $M$ damped harmonic oscillator modes reads:\cite{Barker1975:4071} 
\begin{equation}\label{gLST}
	\sum\limits_i^N \frac{\tau_{Ti}}{\tau_{Li}} \sum\limits_j^M \frac{(\omega^2_{Lj}+\gamma^2_{Lj})}{\omega^2_{Tj}} = \frac{\varepsilon(0)}{\varepsilon_\infty}
\end{equation}

Barker's exposition on the gLST equation suggests it can help distinguish the validity of various fit functions, for instance, whether one should fit with a single Debye relaxation or two overlapping Debye relaxations.\cite{Barker1975:4071} However, to do this, one needs to work with the entire spectrum, so that all the modes that contribute to $\varepsilon(0)$ can be accounted for.

We attempt to do this by first rearranging the dielectric function given in eqn.\ \ref{generalepsomega} into the following form: 
\begin{equation}\label{symformeps}
    \varepsilon(\omega) = \varepsilon_\infty \prod\limits_i^N \frac{(\omega - i \omega_{Li}) }{ (\omega - i\omega_{Ti}) }     \prod\limits_j^M \frac{\omega_{Lj}^2 - \omega^2 - i\omega\gamma_j  }{\omega_{Tj}^2 - \omega^2 - i\omega\gamma_j }  
\end{equation}
The longitudinal frequencies correspond to points where $\varepsilon(\omega) = 0$. Solving for the longitudinal frequencies requires factoring a messy $(N + 2M)$th degree polynomial, which can be done numerically. An easier alternative method is to simply invert eqn.\ \ref{symformeps}
\begin{equation}
  \frac{1}{ \varepsilon(\omega)} = \frac{1}{\varepsilon_\infty} \prod\limits_i^N \frac{(\omega - i \omega_{Ti}) }{ (\omega -i\omega_{Li}) }     \prod\limits_j^M \frac{\omega_{Tj}^2 - \omega^2 - i\omega\gamma_j  }{\omega_{Lj}^2 - \omega^2 - i\omega\gamma_j }  
\end{equation}
By symmetry it should be easy to see that the dielectric function for $\frac{1}{\varepsilon(\omega)}$ has the same form as for $\varepsilon(\omega)$, but with the transverse frequencies changed to the longitudinal frequencies. Thus, we can do a separate fit of $\frac{1}{\varepsilon(\omega)}$ using the same dielectric function given in eqn.\ \ref{generalepsomega} (with the change of $\varepsilon_\infty \rightarrow \frac{1}{\varepsilon_\infty}$) to obtain the longitudinal frequencies. 

This method of fitting allows us to try fitting with the gLST equation as a novel constraint. Alternatively, one could use the Kramers-Kronig relations directly as a constraint, but this requires one has the entire spectrum available. Sometimes the optimization process is not able to satisfy the gLST constraint. Since the \textit{spectrumfitter} code shows the contribution to the left hand side of gLST equation (eqn.\ \ref{gLST}) from each lineshape, one can often pinpoint features in the spectrum are problematic to meeting the constraint. 


\subsection{Results of fitting $f$-sum and gLST constraints}
\begin{figure}[h]
    \centering
        \includegraphics[width=7.2cm]{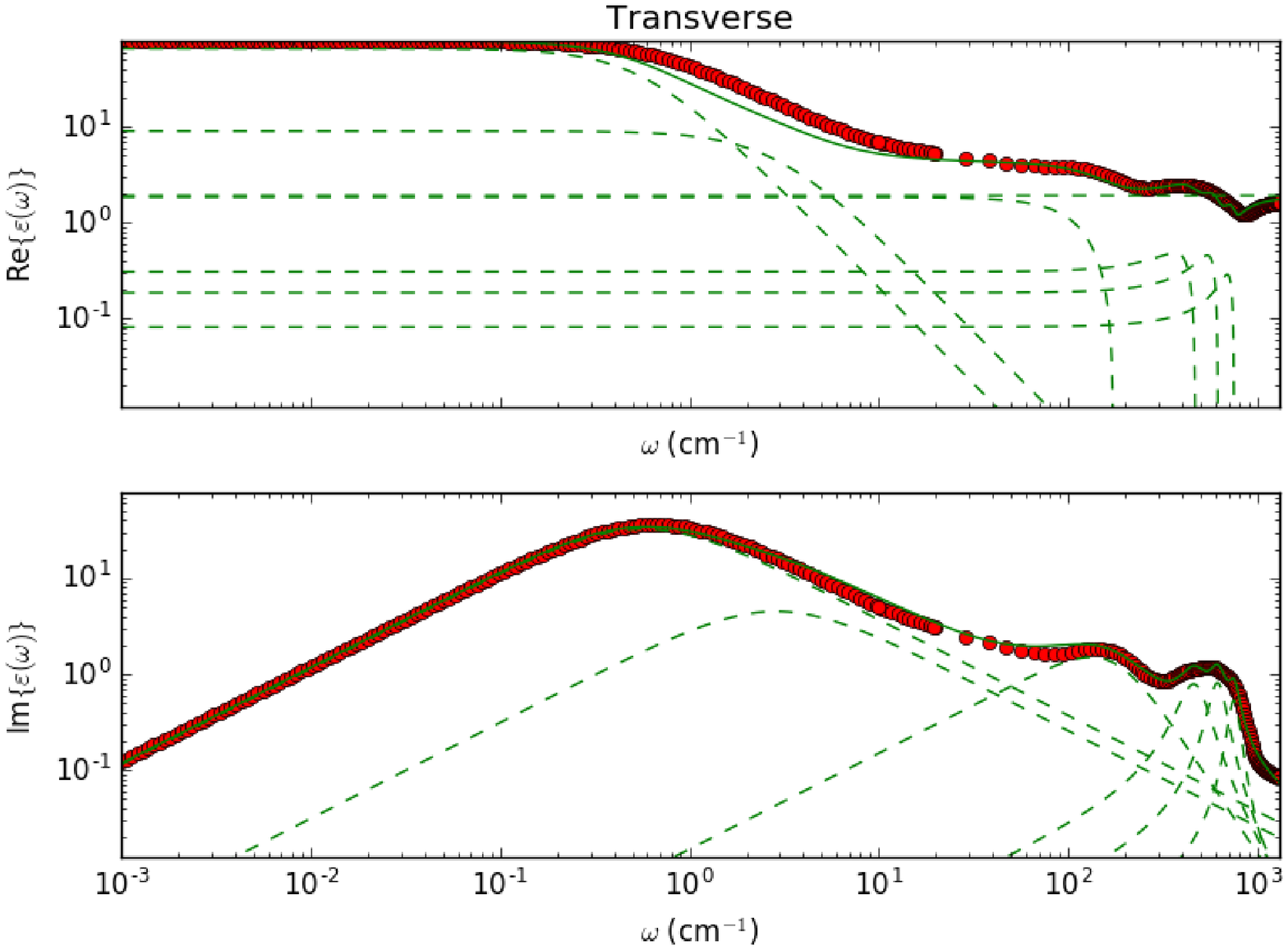}
        \includegraphics[width=7.8cm]{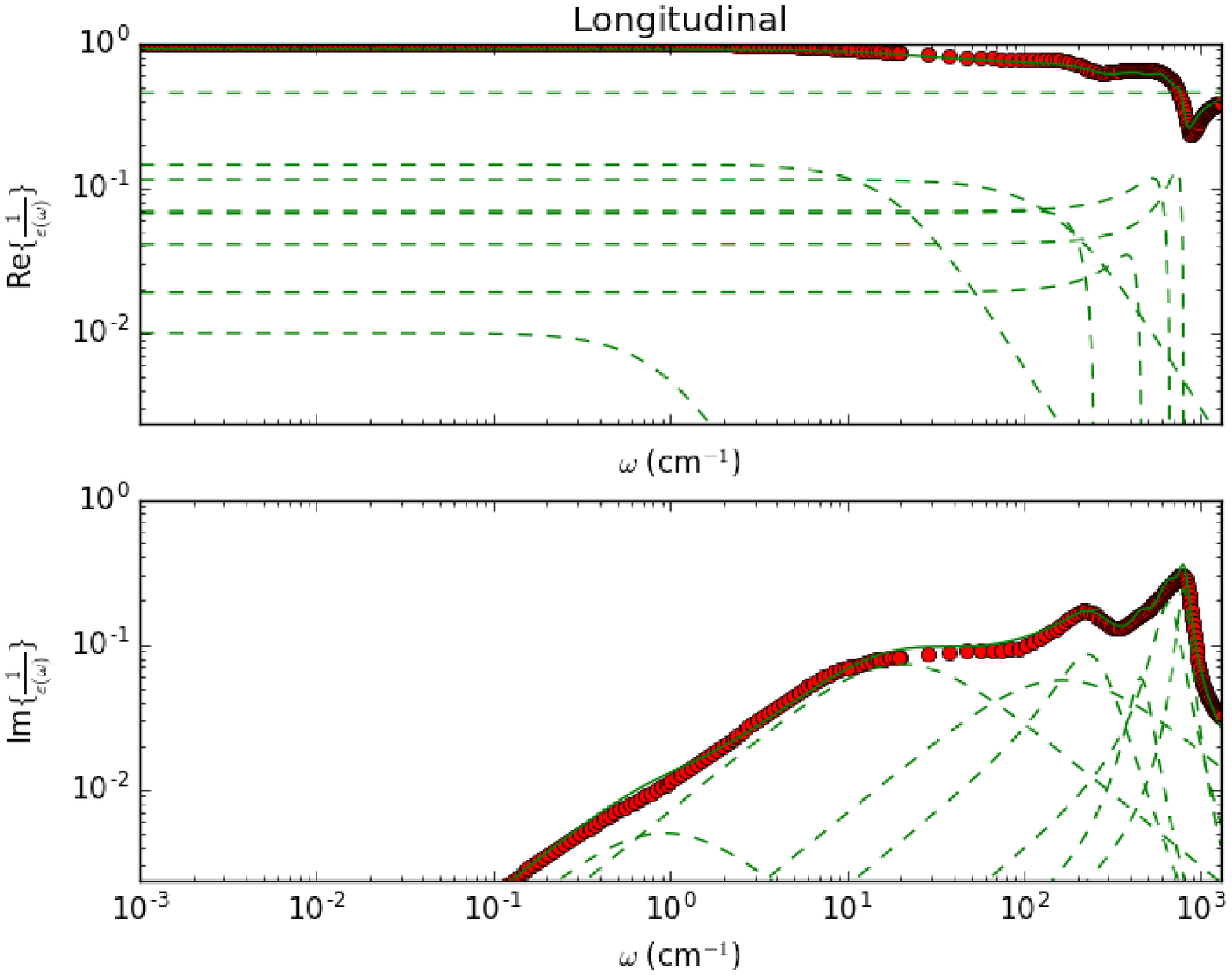}
     \begin{tabular}{c c c c c c}
name (transverse) & $f$ & $\omega_0$ & $\tau$ (ps) &  $\gamma$   \\ 
\hline
               Debye & 65.0 &   0.57 &  9.32 &    \\
           2nd Debye & 8.97 &   2.86 &  1.85 &    \\
           3rd Debye & 0.00 &  39.25 &  0.14 \\
         H-bond str. & 1.85 &  172 &  0.031 & 242  \\
                  L1 & 0.32 &  467 &  0.011 & 184  \\
                  L2 & 0.19 &  616 &  0.009 & 141  \\
                  L3 & 0.08 &  750 &  0.007 & 115  \\
             eps inf & 1.91 & & & & \\
name (longitudinal)   & $f$ & $\omega_0$& $\tau$ (ps) &  $\gamma$  \\ 
\hline
               Debye & 0.01 &   0.93 &  5.70 &    \\
           2nd Debye & 0.15 &  20.4 &  0.26 &    \\
           3rd Debye & 0.11 & 165. &  0.03 &    \\
           Hbond Str & 0.07 &  250 &  0.021 & 203   \\
                  L1 & 0.02 &  463 &  0.011 & 151  \\
                  L2 & 0.07 &  665 &  0.008 & 240   \\
                  L3 & 0.04 &  800 &  0.007 & 135   \\
           eps inf L & 0.46 & & & & \\
     \end{tabular}
     \caption{Example transverse (top) and longitudinal (bottom) fits. $\omega_0$ and $\gamma$ are all reported in cm$^{-1}$. The fit contained 3 Debye relaxations, 1 DHO peak for H-bond stretching and 3 DHO peaks for the librational region. The RMS error was 0.124.}
     \label{gLSTfitDHO}
\end{figure}

We performed fitting with the $f$-sum rule and gLST equation as constraints. The inclusion of the gLST equation as a constraint did not significantly improve the quality of fit, but it did yield some insights into the physicality of the fit functions being used. Figure \ref{gLSTfitDHO} shows a fit with 3 Debye relaxations, 1 DHO peak for H-bond stretching and 3 DHO peaks for the librational region. Under this model, the LHS of the gLST relation here is 284 while the RHS is 48.15. The 2nd Debye relaxation contributes a factor of 7.11 and the 3rd Debye relaxation contributes a factor of 4.2 to the gLST relation. Next we gave the fitting fitting procedure the option of suppressing the 3rd Debye relaxation. It was completely suppressed in the transverse case ($f \rightarrow 0$), suggesting that it is not physical. We found that if we tried to include an additional DHO lineshape  between 10 and 100 cm$^{-1}$ for the H-bond bending peak it was also completely suppressed by the fitting procedure. 

We found a better fit can be obtained using the ``Brendel'' lineshape instead of a damped harmonic oscillator for the librational and stretching peaks. The Brendel lineshape is a convolution of the DHO lineshape with Gaussian:\cite{Brendel1992:1}
\begin{equation}
    \varepsilon(\omega) = \frac{1}{\sqrt{2\pi}\sigma}\int_{-\infty}^{\infty} e^{-\frac{(x - \omega_0)^2}{2\sigma^2}} \frac{ \omega_p^2 }{ x^2 - \omega^2 + i\omega\gamma} + \varepsilon_\infty 
\end{equation}
This lineshape can be calculated analytically in terms of complex error functions, making it easy to evaluate numerically.\cite{Brendel1992:1} We found the Brendel lineshape especially helpful for fitting the H-bond stretching peak at $\approx 150$ cm$^{-1}$. Figure \ref{gLSTfit} shows a fit with 3 Debye relaxations, 2 Brendel peaks for H-bond stretching and H-bond bending, and 3 Brendel peaks for the librational region. To perform the gLST analysis with the Brendel peaks, we consider the Brendel lineshape can be approximately factored into the gLST equation as a single DHO. Then, the LHS of the gLST relation is 456 while the RHS is 48.15. The 1st Debye relaxation contributes a factor of 1.55 to the gLST relation, while the 2nd Debye relaxation contributes a factor of 5.49 and the 3rd Debye relaxation contributes a factor of 6.48. 

In every type of fit we tried, inclusion of the 3rd Debye relaxation resulted in a significantly larger departure from the the gLST equation. If only 2 Debye relaxations are used in this model, the gLST LHS becomes 21.921, while the RMS error increases slightly from 0.120 to 0.125. 

\section{Conclusion}
We critically reviewed the literature on the Debye relaxation and the high frequency excess response and found widespread disagreement on the molecular mechanisms underlying these processes. We studied the distance decomposition of the Debye relaxation, finding that it involves significant correlations between molecules on scales of 1.5 - 2 nm. The $k$-dependence of Debye relaxation suggests that Debye relaxation is a collective process, and the dispersion relation of the Debye peak suggests a propagating mode. Our findings call into question models of Debye relaxation such as Agmon's model,\cite{Agmon1996:1072} which related Debye relaxation to rotation/translation after a hydrogen bond breaks, and Buchner's model, which related it to the rotation of nearly free molecules.\cite{Buchner1999:57} 

Using our \textit{spectrumfitter} Python package we attempted fitting experimental spectra with secondary and tertiary Debye processes to model the high frequency excess. When we tried to apply the gLST relation to our fitting we found that both the fits with one secondary and an additional tertiary Debye process are problematic. When the gLST relation is used as a constraint with such fits, large violations must be incurred, especially when the tertiary Debye process is included. Given the understanding of H-bond network dynamics that has been elucidated by the work of many authors during the last few decades\cite{Sciortino1990:3452,Ohmine1995:6767,Heyden2010:12068,Kindt1996:10373} as well as in our recent work,\cite{Elton2016:10193} we believe there is a wide distribution of H-bond network modes which contribute in this region. In light of this, the lack of agreement of values for $\tau_2$ and $\tau_3$ reported in the previous literature (table I) is not surprising. The narrow frequency windows of most experiments and failure to unbias when fitting further contribute to the dispersion in values. 

The totality of our findings are consistent with the recent model of Debye relaxation by Popov et al.,\cite{Popov2016:13941} which posits that it is caused by the movement of defects in the hydrogen bond network. These defects are similar to the L and D Bjerrum defects found in ice, but likely more smeared out.\cite{Popov2016:13941} The existence of such defects in liquid water, under the name of ``bifuricated'' H-bonds has already been established from x-ray scattering\cite{Sciortino1990:3452} and simulation.\cite{Laage2006:832} The migration of these defects through the H-bond network explains the semi-long range dependence we found ($\approx$ 15 $\Ang$) as well as the dispersion of the Debye peak we found from simulation and that was found in X-ray scattering experiments.\cite{Arbe2016:185501} The migration of these defects is not purely diffusive, but occurs due to hopping motions. The idea is attractive in light of our recent work showing ice-like phonons that propagate through the H-bond network.\cite{Elton2016:10193} If the dynamics of water are ice-like at the librational frequencies $400-800$ cm$^{-1}$, then it is not surprising they are also ice-like at lower frequencies. Dielectric relaxation in ice is almost perfectly described by a single Debye relaxation, at least at high temperatures.\cite{hobbs2010ice} 

According to the theory of Popov et al., the hopping motion of a defect results in both translational and oscillatory motions of water molecules, which cause vibrations in the H-bond network. Under this assumption, and taking into account the power-law vibrational dynamics of the H-bond network, Popov et al.\ are able to derive the excess wing of the Debye relaxation.\cite{Popov2016:13941} Thus we also see that the theory of Popov et al.\ helps resolve the mystery of how a single Debye peak arises from the heterogeneous (and fractal-like) H-bond network. The heterogeneous and fractal-like nature of the H-bond network does in fact appear in the spectrum, giving rise to the excess response. In other words, the Debye peak in liquid water is not a standard pure Debye peak, but is a Debye peak whose high frequency side is modified by the presence of the complex H-bond network.  


\section{Acknowledgements}
This work was supported by DOE Award No.\ DE-FG02-09ER16052. The author would like to thank his Ph.D.\ adviser, Prof.\ Marivi Fern\'{a}ndez-Serra for allowing him to pursue this research and providing guidance and critical proofreading of the manuscript. The author also acknowledges Dr.\ Kevin Hauser, Dr.\ Ivan Popov, Prof.\ Marcello Sega, and Prof.\ John Watterson for providing useful feedback. 

\bibliographystyle{apsrev}
\bibliography{combined_refs.bib}

\begin{thebibliography}{95}
\expandafter\ifx\csname natexlab\endcsname\relax\def\natexlab#1{#1}\fi
\expandafter\ifx\csname bibnamefont\endcsname\relax
  \def\bibnamefont#1{#1}\fi
\expandafter\ifx\csname bibfnamefont\endcsname\relax
  \def\bibfnamefont#1{#1}\fi
\expandafter\ifx\csname citenamefont\endcsname\relax
  \def\citenamefont#1{#1}\fi
\expandafter\ifx\csname url\endcsname\relax
  \def\url#1{\texttt{#1}}\fi
\expandafter\ifx\csname urlprefix\endcsname\relax\def\urlprefix{URL }\fi
\providecommand{\bibinfo}[2]{#2}
\providecommand{\eprint}[2][]{\url{#2}}

\bibitem[{\citenamefont{Rosenkranz}(2015)}]{Rosenkranz2015:1387}
\bibinfo{author}{\bibfnamefont{P.}~\bibnamefont{Rosenkranz}},
  \bibinfo{journal}{Geoscience and Remote Sensing, IEEE Transactions on}
  \textbf{\bibinfo{volume}{53}}, \bibinfo{pages}{1387} (\bibinfo{year}{2015}).

\bibitem[{\citenamefont{Wang and Schmugge}(1980)}]{Wang1980:288}
\bibinfo{author}{\bibfnamefont{J.}~\bibnamefont{Wang}} \bibnamefont{and}
  \bibinfo{author}{\bibfnamefont{T.}~\bibnamefont{Schmugge}},
  \bibinfo{journal}{Geoscience and Remote Sensing, IEEE Transactions on}
  \textbf{\bibinfo{volume}{GE-18}}, \bibinfo{pages}{288}
  (\bibinfo{year}{1980}).

\bibitem[{\citenamefont{Kneifel et~al.}(2014)\citenamefont{Kneifel, Redl,
  Orlandi, L\"{o}hnert, Cadeddu, Turner, and Chen}}]{Kneifel2014}
\bibinfo{author}{\bibfnamefont{S.}~\bibnamefont{Kneifel}},
  \bibinfo{author}{\bibfnamefont{S.}~\bibnamefont{Redl}},
  \bibinfo{author}{\bibfnamefont{E.}~\bibnamefont{Orlandi}},
  \bibinfo{author}{\bibfnamefont{U.}~\bibnamefont{L\"{o}hnert}},
  \bibinfo{author}{\bibfnamefont{M.}~\bibnamefont{Cadeddu}},
  \bibinfo{author}{\bibfnamefont{D.}~\bibnamefont{Turner}}, \bibnamefont{and}
  \bibinfo{author}{\bibfnamefont{M.}~\bibnamefont{Chen}},
  \bibinfo{journal}{journal of Applied Meteorology and Climatology}
  \textbf{\bibinfo{volume}{53}}, \bibinfo{pages}{1028} (\bibinfo{year}{2014}).

\bibitem[{\citenamefont{Nilsson and Pettersson}(2016)}]{Nilsson2016:8988}
\bibinfo{author}{\bibfnamefont{A.}~\bibnamefont{Nilsson}} \bibnamefont{and}
  \bibinfo{author}{\bibfnamefont{L.~G.~M.} \bibnamefont{Pettersson}},
  \bibinfo{journal}{Nat. Comm.} \textbf{\bibinfo{volume}{6}},
  \bibinfo{pages}{8988} (\bibinfo{year}{2016}).

\bibitem[{\citenamefont{Limmer and Chandler}(2011)}]{Limmer2011:134503}
\bibinfo{author}{\bibfnamefont{D.~T.} \bibnamefont{Limmer}} \bibnamefont{and}
  \bibinfo{author}{\bibfnamefont{D.}~\bibnamefont{Chandler}},
  \bibinfo{journal}{J. Chem. Phys.} \textbf{\bibinfo{volume}{135}},
  \bibinfo{eid}{134503} (\bibinfo{year}{2011}).

\bibitem[{\citenamefont{Ceriotti et~al.}(2016)\citenamefont{Ceriotti, Fang,
  Kusalik, McKenzie, Michaelides, Morales, and Markland}}]{Ceriotti2016review}
\bibinfo{author}{\bibfnamefont{M.}~\bibnamefont{Ceriotti}},
  \bibinfo{author}{\bibfnamefont{W.}~\bibnamefont{Fang}},
  \bibinfo{author}{\bibfnamefont{P.~G.} \bibnamefont{Kusalik}},
  \bibinfo{author}{\bibfnamefont{R.~H.} \bibnamefont{McKenzie}},
  \bibinfo{author}{\bibfnamefont{A.}~\bibnamefont{Michaelides}},
  \bibinfo{author}{\bibfnamefont{M.~A.} \bibnamefont{Morales}},
  \bibnamefont{and} \bibinfo{author}{\bibfnamefont{T.~E.}
  \bibnamefont{Markland}}, \bibinfo{journal}{Chem. Rev.}
  \textbf{\bibinfo{volume}{116}}, \bibinfo{pages}{7529} (\bibinfo{year}{2016}).

\bibitem[{\citenamefont{Popov et~al.}(2016)\citenamefont{Popov, Ishai, Khamzin,
  and Feldman}}]{Popov2016:13941}
\bibinfo{author}{\bibfnamefont{I.}~\bibnamefont{Popov}},
  \bibinfo{author}{\bibfnamefont{P.~B.} \bibnamefont{Ishai}},
  \bibinfo{author}{\bibfnamefont{A.}~\bibnamefont{Khamzin}}, \bibnamefont{and}
  \bibinfo{author}{\bibfnamefont{Y.}~\bibnamefont{Feldman}},
  \bibinfo{journal}{Phys. Chem. Chem. Phys.} \textbf{\bibinfo{volume}{18}},
  \bibinfo{pages}{13941} (\bibinfo{year}{2016}).

\bibitem[{\citenamefont{Ben~Ishai et~al.}(2015)\citenamefont{Ben~Ishai,
  Tripathi, Kawase, Puzenko, and Feldman}}]{Ishai2015:15428}
\bibinfo{author}{\bibfnamefont{P.}~\bibnamefont{Ben~Ishai}},
  \bibinfo{author}{\bibfnamefont{S.~R.} \bibnamefont{Tripathi}},
  \bibinfo{author}{\bibfnamefont{K.}~\bibnamefont{Kawase}},
  \bibinfo{author}{\bibfnamefont{A.}~\bibnamefont{Puzenko}}, \bibnamefont{and}
  \bibinfo{author}{\bibfnamefont{Y.}~\bibnamefont{Feldman}},
  \bibinfo{journal}{Phys. Chem. Chem. Phys.} \textbf{\bibinfo{volume}{17}},
  \bibinfo{pages}{15428} (\bibinfo{year}{2015}).

\bibitem[{\citenamefont{Hansen et~al.}(2016)\citenamefont{Hansen, Kisliuk,
  Sokolov, and Gainaru}}]{Hansen2016:237601}
\bibinfo{author}{\bibfnamefont{J.~S.} \bibnamefont{Hansen}},
  \bibinfo{author}{\bibfnamefont{A.}~\bibnamefont{Kisliuk}},
  \bibinfo{author}{\bibfnamefont{A.~P.} \bibnamefont{Sokolov}},
  \bibnamefont{and} \bibinfo{author}{\bibfnamefont{C.}~\bibnamefont{Gainaru}},
  \bibinfo{journal}{Phys. Rev. Lett.} \textbf{\bibinfo{volume}{116}},
  \bibinfo{pages}{237601} (\bibinfo{year}{2016}).

\bibitem[{\citenamefont{Arbe et~al.}(2016)\citenamefont{Arbe, Malo~de Molina,
  Alvarez, Frick, and Colmenero}}]{Arbe2016:185501}
\bibinfo{author}{\bibfnamefont{A.}~\bibnamefont{Arbe}},
  \bibinfo{author}{\bibfnamefont{P.}~\bibnamefont{Malo~de Molina}},
  \bibinfo{author}{\bibfnamefont{F.}~\bibnamefont{Alvarez}},
  \bibinfo{author}{\bibfnamefont{B.}~\bibnamefont{Frick}}, \bibnamefont{and}
  \bibinfo{author}{\bibfnamefont{J.}~\bibnamefont{Colmenero}},
  \bibinfo{journal}{Phys. Rev. Lett.} \textbf{\bibinfo{volume}{117}},
  \bibinfo{pages}{185501} (\bibinfo{year}{2016}).

\bibitem[{\citenamefont{Lunkenheimer et~al.}(2016)\citenamefont{Lunkenheimer,
  Emmert, Gulich, Köhler, Wolf, Schwab, and Loidl}}]{Lunkenheimer:1612.01457}
\bibinfo{author}{\bibfnamefont{P.}~\bibnamefont{Lunkenheimer}},
  \bibinfo{author}{\bibfnamefont{S.}~\bibnamefont{Emmert}},
  \bibinfo{author}{\bibfnamefont{R.}~\bibnamefont{Gulich}},
  \bibinfo{author}{\bibfnamefont{M.}~\bibnamefont{Köhler}},
  \bibinfo{author}{\bibfnamefont{M.}~\bibnamefont{Wolf}},
  \bibinfo{author}{\bibfnamefont{M.}~\bibnamefont{Schwab}}, \bibnamefont{and}
  \bibinfo{author}{\bibfnamefont{A.}~\bibnamefont{Loidl}},
  \emph{\bibinfo{title}{Dielectric relaxation dynamics and the boson peak of
  water}} (\bibinfo{year}{2016}), \eprint{arXiv:1612.01457}.

\bibitem[{\citenamefont{Ronne et~al.}(1997)\citenamefont{Ronne, Thrane,
  Åstrand, Wallqvist, Mikkelsen, and Keiding}}]{Ronne1997:5319}
\bibinfo{author}{\bibfnamefont{C.}~\bibnamefont{Ronne}},
  \bibinfo{author}{\bibfnamefont{L.}~\bibnamefont{Thrane}},
  \bibinfo{author}{\bibfnamefont{P.-O.} \bibnamefont{Åstrand}},
  \bibinfo{author}{\bibfnamefont{A.}~\bibnamefont{Wallqvist}},
  \bibinfo{author}{\bibfnamefont{K.~V.} \bibnamefont{Mikkelsen}},
  \bibnamefont{and} \bibinfo{author}{\bibfnamefont{S.~R.}
  \bibnamefont{Keiding}}, \bibinfo{journal}{J. Chem. Phys.}
  \textbf{\bibinfo{volume}{107}} (\bibinfo{year}{1997}).

\bibitem[{Vij(2004)}]{Vij2004125}
\bibinfo{journal}{J. Mol. Liq.} \textbf{\bibinfo{volume}{112}},
  \bibinfo{pages}{125 } (\bibinfo{year}{2004}).

\bibitem[{\citenamefont{Agmon}(1996)}]{Agmon1996:1072}
\bibinfo{author}{\bibfnamefont{N.}~\bibnamefont{Agmon}}, \bibinfo{journal}{J.
  Phys. Chem.} \textbf{\bibinfo{volume}{100}}, \bibinfo{pages}{1072}
  (\bibinfo{year}{1996}).

\bibitem[{\citenamefont{Buchner et~al.}(1999)\citenamefont{Buchner, Barthel,
  and journal Stauber}}]{Buchner1999:57}
\bibinfo{author}{\bibfnamefont{R.}~\bibnamefont{Buchner}},
  \bibinfo{author}{\bibnamefont{Barthel}}, \bibnamefont{and}
  \bibinfo{author}{\bibnamefont{journal Stauber}}, \bibinfo{journal}{Chem.
  Phys. Lett.} \textbf{\bibinfo{volume}{306}}, \bibinfo{pages}{57}
  (\bibinfo{year}{1999}).

\bibitem[{\citenamefont{Nabokov and Lubimov}(1988)}]{Nabokov:1473}
\bibinfo{author}{\bibfnamefont{O.}~\bibnamefont{Nabokov}} \bibnamefont{and}
  \bibinfo{author}{\bibfnamefont{Y.}~\bibnamefont{Lubimov}},
  \bibinfo{journal}{Mol. Phys.} \textbf{\bibinfo{volume}{65}},
  \bibinfo{pages}{1473} (\bibinfo{year}{1988}).

\bibitem[{\citenamefont{Stanley and Teixeira}(1980)}]{stanley:3404}
\bibinfo{author}{\bibfnamefont{H.~E.} \bibnamefont{Stanley}} \bibnamefont{and}
  \bibinfo{author}{\bibfnamefont{J.}~\bibnamefont{Teixeira}},
  \bibinfo{journal}{The Journal of Chemical Physics}
  \textbf{\bibinfo{volume}{73}}, \bibinfo{pages}{3404} (\bibinfo{year}{1980}).

\bibitem[{\citenamefont{J.B.~Hasted}(1985)}]{Hasted1985:622}
\bibinfo{author}{\bibfnamefont{F.~F. J.~B.} \bibnamefont{J.B.~Hasted},
  \bibfnamefont{S.K.~Husain}}, \bibinfo{journal}{Chem. Phys. Lett.}
  \textbf{\bibinfo{volume}{118}}, \bibinfo{pages}{622} (\bibinfo{year}{1985}).

\bibitem[{\citenamefont{Zasetsky and Buchner}(2011)}]{Zasetsky2011:025903}
\bibinfo{author}{\bibfnamefont{A.~Y.} \bibnamefont{Zasetsky}} \bibnamefont{and}
  \bibinfo{author}{\bibfnamefont{R.}~\bibnamefont{Buchner}},
  \bibinfo{journal}{journal of Physics: Condensed Matter}
  \textbf{\bibinfo{volume}{23}}, \bibinfo{pages}{025903}
  (\bibinfo{year}{2011}).

\bibitem[{\citenamefont{von Hippel}(1988)}]{vonHippel1988}
\bibinfo{author}{\bibfnamefont{A.~R.} \bibnamefont{von Hippel}},
  \bibinfo{journal}{IEEE Transactions on Electrical Insulation}
  \textbf{\bibinfo{volume}{23}}, \bibinfo{pages}{825} (\bibinfo{year}{1988}).

\bibitem[{\citenamefont{Arkhipov}(2002)}]{Arkhipov:127}
\bibinfo{author}{\bibfnamefont{V.}~\bibnamefont{Arkhipov}},
  \bibinfo{journal}{journal of Non-Crystalline Solids}
  \textbf{\bibinfo{volume}{305}}, \bibinfo{pages}{127 } (\bibinfo{year}{2002}).

\bibitem[{\citenamefont{Arkhipov and Agmon}(2003)}]{Arkhipov2003}
\bibinfo{author}{\bibfnamefont{V.~I.} \bibnamefont{Arkhipov}} \bibnamefont{and}
  \bibinfo{author}{\bibfnamefont{N.}~\bibnamefont{Agmon}},
  \bibinfo{journal}{Israel Journal of Chemistry} \textbf{\bibinfo{volume}{43}},
  \bibinfo{pages}{363} (\bibinfo{year}{2003}), ISSN \bibinfo{issn}{1869-5868},
  \urlprefix\url{http://dx.doi.org/10.1560/5WKJ-WJ9F-Q0DR-WPFH}.

\bibitem[{\citenamefont{Artemov and Volkov}(2014)}]{Artemov2014:158}
\bibinfo{author}{\bibfnamefont{V.~G.} \bibnamefont{Artemov}} \bibnamefont{and}
  \bibinfo{author}{\bibfnamefont{A.~A.} \bibnamefont{Volkov}},
  \bibinfo{journal}{Ferroelectrics} \textbf{\bibinfo{volume}{466}},
  \bibinfo{pages}{158} (\bibinfo{year}{2014}).

\bibitem[{\citenamefont{Hobbs}(2010)}]{hobbs2010ice}
\bibinfo{author}{\bibfnamefont{P.}~\bibnamefont{Hobbs}},
  \emph{\bibinfo{title}{Ice Physics}}, Oxford Classic Texts in the Physical
  Sciences (\bibinfo{year}{2010}), ISBN \bibinfo{isbn}{9780199587711}.

\bibitem[{\citenamefont{Kindt and Schmuttenmaer}(1996)}]{Kindt1996:10373}
\bibinfo{author}{\bibfnamefont{J.~T.} \bibnamefont{Kindt}} \bibnamefont{and}
  \bibinfo{author}{\bibfnamefont{C.~A.} \bibnamefont{Schmuttenmaer}},
  \bibinfo{journal}{J. Phys. Chem.} \textbf{\bibinfo{volume}{100}},
  \bibinfo{pages}{10373} (\bibinfo{year}{1996}).

\bibitem[{\citenamefont{Kaatze}(1993)}]{Kaatze1993:95}
\bibinfo{author}{\bibfnamefont{U.}~\bibnamefont{Kaatze}}, \bibinfo{journal}{J.
  Mol. Liq.} \textbf{\bibinfo{volume}{56}}, \bibinfo{pages}{95 }
  (\bibinfo{year}{1993}).

\bibitem[{\citenamefont{Mason et~al.}(1974)\citenamefont{Mason, Hasted, and
  Moore}}]{Mason1974217}
\bibinfo{author}{\bibfnamefont{P.}~\bibnamefont{Mason}},
  \bibinfo{author}{\bibfnamefont{J.}~\bibnamefont{Hasted}}, \bibnamefont{and}
  \bibinfo{author}{\bibfnamefont{L.}~\bibnamefont{Moore}},
  \bibinfo{journal}{Advances in Molecular Relaxation Processes}
  \textbf{\bibinfo{volume}{6}}, \bibinfo{pages}{217 } (\bibinfo{year}{1974}).

\bibitem[{\citenamefont{Saito and Ohmine}(1994)}]{Saito1994:6063}
\bibinfo{author}{\bibfnamefont{S.}~\bibnamefont{Saito}} \bibnamefont{and}
  \bibinfo{author}{\bibfnamefont{I.}~\bibnamefont{Ohmine}},
  \bibinfo{journal}{J. Chem. Phys.} \textbf{\bibinfo{volume}{101}},
  \bibinfo{pages}{6063} (\bibinfo{year}{1994}).

\bibitem[{\citenamefont{Baba et~al.}(1997)\citenamefont{Baba, Hirata, Saito,
  Ohmine, and Wales}}]{babaSaito1997:3329}
\bibinfo{author}{\bibfnamefont{A.}~\bibnamefont{Baba}},
  \bibinfo{author}{\bibfnamefont{Y.}~\bibnamefont{Hirata}},
  \bibinfo{author}{\bibfnamefont{S.}~\bibnamefont{Saito}},
  \bibinfo{author}{\bibfnamefont{I.}~\bibnamefont{Ohmine}}, \bibnamefont{and}
  \bibinfo{author}{\bibfnamefont{D.~J.} \bibnamefont{Wales}},
  \bibinfo{journal}{J. Chem. Phys.} \textbf{\bibinfo{volume}{106}},
  \bibinfo{pages}{3329} (\bibinfo{year}{1997}).

\bibitem[{\citenamefont{Luzar and Chandler}(1996)}]{Luzar1996:55}
\bibinfo{author}{\bibfnamefont{A.}~\bibnamefont{Luzar}} \bibnamefont{and}
  \bibinfo{author}{\bibfnamefont{D.}~\bibnamefont{Chandler}},
  \bibinfo{journal}{Nature} \textbf{\bibinfo{volume}{379}}, \bibinfo{pages}{55}
  (\bibinfo{year}{1996}).

\bibitem[{\citenamefont{Kumar et~al.}(2006)\citenamefont{Kumar, Franzese,
  Buldyrev, and Stanley}}]{Kumar:041505}
\bibinfo{author}{\bibfnamefont{P.}~\bibnamefont{Kumar}},
  \bibinfo{author}{\bibfnamefont{G.}~\bibnamefont{Franzese}},
  \bibinfo{author}{\bibfnamefont{S.~V.} \bibnamefont{Buldyrev}},
  \bibnamefont{and} \bibinfo{author}{\bibfnamefont{H.~E.}
  \bibnamefont{Stanley}}, \bibinfo{journal}{Phys. Rev. E}
  \textbf{\bibinfo{volume}{73}}, \bibinfo{pages}{041505}
  (\bibinfo{year}{2006}).

\bibitem[{\citenamefont{Ohmine and Tanaka}(1993)}]{Ohmine1993:2545}
\bibinfo{author}{\bibfnamefont{I.}~\bibnamefont{Ohmine}} \bibnamefont{and}
  \bibinfo{author}{\bibfnamefont{H.}~\bibnamefont{Tanaka}},
  \bibinfo{journal}{Chem. Rev.} \textbf{\bibinfo{volume}{93}},
  \bibinfo{pages}{2545} (\bibinfo{year}{1993}).

\bibitem[{\citenamefont{Shiratani and Sasai}(1996)}]{Shiratani1996:7671}
\bibinfo{author}{\bibfnamefont{E.}~\bibnamefont{Shiratani}} \bibnamefont{and}
  \bibinfo{author}{\bibfnamefont{M.}~\bibnamefont{Sasai}}, \bibinfo{journal}{J.
  Chem. Phys.} \textbf{\bibinfo{volume}{104}} (\bibinfo{year}{1996}).

\bibitem[{\citenamefont{Yamaguchi et~al.}(2003)\citenamefont{Yamaguchi, Chong,
  and Hirata}}]{Yamaguchi2003:1211}
\bibinfo{author}{\bibfnamefont{T.}~\bibnamefont{Yamaguchi}},
  \bibinfo{author}{\bibfnamefont{S.-H.} \bibnamefont{Chong}}, \bibnamefont{and}
  \bibinfo{author}{\bibfnamefont{F.}~\bibnamefont{Hirata}},
  \bibinfo{journal}{Mol. Phys.} \textbf{\bibinfo{volume}{101}},
  \bibinfo{pages}{1211} (\bibinfo{year}{2003}).

\bibitem[{\citenamefont{Yonetani}(2005)}]{Yonetani:49}
\bibinfo{author}{\bibfnamefont{Y.}~\bibnamefont{Yonetani}},
  \bibinfo{journal}{Chem. Phys. Lett.} \textbf{\bibinfo{volume}{406}},
  \bibinfo{pages}{49 } (\bibinfo{year}{2005}).

\bibitem[{\citenamefont{van~der Spoel and van Maaren}(2006)}]{Spoel:1}
\bibinfo{author}{\bibfnamefont{D.}~\bibnamefont{van~der Spoel}}
  \bibnamefont{and} \bibinfo{author}{\bibfnamefont{P.~J.} \bibnamefont{van
  Maaren}}, \bibinfo{journal}{J. Chem. Theo. Comp.}
  \textbf{\bibinfo{volume}{2}}, \bibinfo{pages}{1} (\bibinfo{year}{2006}).

\bibitem[{\citenamefont{Vinh et~al.}(2015)\citenamefont{Vinh, Sherwin, Allen,
  George, Rahmani, and Plaxco}}]{Vinh2015:164502}
\bibinfo{author}{\bibfnamefont{N.~Q.} \bibnamefont{Vinh}},
  \bibinfo{author}{\bibfnamefont{M.~S.} \bibnamefont{Sherwin}},
  \bibinfo{author}{\bibfnamefont{S.~J.} \bibnamefont{Allen}},
  \bibinfo{author}{\bibfnamefont{D.~K.} \bibnamefont{George}},
  \bibinfo{author}{\bibfnamefont{A.~J.} \bibnamefont{Rahmani}},
  \bibnamefont{and} \bibinfo{author}{\bibfnamefont{K.~W.}
  \bibnamefont{Plaxco}}, \bibinfo{journal}{J. Chem. Phys.}
  \textbf{\bibinfo{volume}{142}}, \bibinfo{pages}{164502}
  (\bibinfo{year}{2015}).

\bibitem[{\citenamefont{Turton et~al.}(2008)\citenamefont{Turton, Hunger,
  Hefter, Buchner, and Wynne}}]{Turton}
\bibinfo{author}{\bibfnamefont{D.~A.} \bibnamefont{Turton}},
  \bibinfo{author}{\bibfnamefont{J.}~\bibnamefont{Hunger}},
  \bibinfo{author}{\bibfnamefont{G.}~\bibnamefont{Hefter}},
  \bibinfo{author}{\bibfnamefont{R.}~\bibnamefont{Buchner}}, \bibnamefont{and}
  \bibinfo{author}{\bibfnamefont{K.}~\bibnamefont{Wynne}}, \bibinfo{journal}{J.
  Chem. Phys.} \textbf{\bibinfo{volume}{128}}, \bibinfo{pages}{161102}
  (\bibinfo{year}{2008}).

\bibitem[{\citenamefont{Debye}(1929)}]{D29}
\bibinfo{author}{\bibfnamefont{P.}~\bibnamefont{Debye}},
  \emph{\bibinfo{title}{Polar Molecules}} (\bibinfo{publisher}{Chemical Catalog
  Co.}, \bibinfo{address}{New York}, \bibinfo{year}{1929}).

\bibitem[{\citenamefont{Laage and Hynes}(2008)}]{Laage2008:14230}
\bibinfo{author}{\bibfnamefont{D.}~\bibnamefont{Laage}} \bibnamefont{and}
  \bibinfo{author}{\bibfnamefont{J.~T.} \bibnamefont{Hynes}},
  \bibinfo{journal}{J. Phys. Chem. B} \textbf{\bibinfo{volume}{112}},
  \bibinfo{pages}{14230} (\bibinfo{year}{2008}).

\bibitem[{\citenamefont{Laage and Hynes}(2006)}]{Laage2006:832}
\bibinfo{author}{\bibfnamefont{D.}~\bibnamefont{Laage}} \bibnamefont{and}
  \bibinfo{author}{\bibfnamefont{J.~T.} \bibnamefont{Hynes}},
  \bibinfo{journal}{Science} \textbf{\bibinfo{volume}{311}},
  \bibinfo{pages}{832} (\bibinfo{year}{2006}).

\bibitem[{\citenamefont{Ludwig}(2007)}]{Ludwig2007}
\bibinfo{author}{\bibfnamefont{R.}~\bibnamefont{Ludwig}},
  \bibinfo{journal}{Chem. Phys. Phys. Chem.} \textbf{\bibinfo{volume}{8}},
  \bibinfo{pages}{44} (\bibinfo{year}{2007}), ISSN \bibinfo{issn}{1439-7641}.

\bibitem[{\citenamefont{Geiger}(2003)}]{GEIGER2003131}
\bibinfo{author}{\bibfnamefont{A.}~\bibnamefont{Geiger}},
  \bibinfo{journal}{Journal of Molecular Liquids}
  \textbf{\bibinfo{volume}{106}}, \bibinfo{pages}{131 } (\bibinfo{year}{2003}),
  ISSN \bibinfo{issn}{0167-7322}.

\bibitem[{\citenamefont{Bertolini et~al.}(1982)\citenamefont{Bertolini,
  Cassettari, and Salvetti}}]{bertolini:3285}
\bibinfo{author}{\bibfnamefont{D.}~\bibnamefont{Bertolini}},
  \bibinfo{author}{\bibfnamefont{M.}~\bibnamefont{Cassettari}},
  \bibnamefont{and} \bibinfo{author}{\bibfnamefont{G.}~\bibnamefont{Salvetti}},
  \bibinfo{journal}{J. Chem. Phys.} \textbf{\bibinfo{volume}{76}},
  \bibinfo{pages}{3285} (\bibinfo{year}{1982}).

\bibitem[{\citenamefont{Onsager}(1936)}]{O23}
\bibinfo{author}{\bibfnamefont{J.}~\bibnamefont{Onsager}}, \bibinfo{journal}{J.
  Am. Chem. Soc.} \textbf{\bibinfo{volume}{58}}, \bibinfo{pages}{1486}
  (\bibinfo{year}{1936}).

\bibitem[{\citenamefont{Glarum}(1960)}]{Glarum:1371}
\bibinfo{author}{\bibfnamefont{S.~H.} \bibnamefont{Glarum}},
  \bibinfo{journal}{J. Chem. Phys.} \textbf{\bibinfo{volume}{33}},
  \bibinfo{pages}{1371} (\bibinfo{year}{1960}).

\bibitem[{\citenamefont{Powles}(1953)}]{powles:633}
\bibinfo{author}{\bibfnamefont{J.~G.} \bibnamefont{Powles}},
  \bibinfo{journal}{J. Chem. Phys.} \textbf{\bibinfo{volume}{21}},
  \bibinfo{pages}{633} (\bibinfo{year}{1953}).

\bibitem[{\citenamefont{Bagchi}(2012)}]{BagchiBook2012}
\bibinfo{author}{\bibfnamefont{B.}~\bibnamefont{Bagchi}},
  \emph{\bibinfo{title}{{Molecular Relaxation in Liquids}}}
  (\bibinfo{publisher}{Oxford University Press}, \bibinfo{year}{2012}), ISBN
  \bibinfo{isbn}{0199863326}.

\bibitem[{\citenamefont{Fatuzzo and Mason}(1967)}]{Fatuzzo:729}
\bibinfo{author}{\bibfnamefont{E.}~\bibnamefont{Fatuzzo}} \bibnamefont{and}
  \bibinfo{author}{\bibfnamefont{P.~R.} \bibnamefont{Mason}},
  \bibinfo{journal}{Proceedings of the Physical Society}
  \textbf{\bibinfo{volume}{90}}, \bibinfo{pages}{729} (\bibinfo{year}{1967}).

\bibitem[{\citenamefont{Nee and Zwanzig}(1970)}]{Nee1970:6353}
\bibinfo{author}{\bibfnamefont{T.}~\bibnamefont{Nee}} \bibnamefont{and}
  \bibinfo{author}{\bibfnamefont{R.}~\bibnamefont{Zwanzig}},
  \bibinfo{journal}{J. Chem. Phys.} \textbf{\bibinfo{volume}{52}},
  \bibinfo{pages}{6353} (\bibinfo{year}{1970}).

\bibitem[{\citenamefont{Bagchi and Chandra}(1990)}]{Bagchi1990:455}
\bibinfo{author}{\bibfnamefont{B.}~\bibnamefont{Bagchi}} \bibnamefont{and}
  \bibinfo{author}{\bibfnamefont{A.}~\bibnamefont{Chandra}},
  \bibinfo{journal}{Phys. Rev. Lett.} \textbf{\bibinfo{volume}{64}},
  \bibinfo{pages}{455} (\bibinfo{year}{1990}).

\bibitem[{\citenamefont{Petrenko and Whitworth}(1999)}]{petrenko1999physics}
\bibinfo{author}{\bibfnamefont{V.}~\bibnamefont{Petrenko}} \bibnamefont{and}
  \bibinfo{author}{\bibfnamefont{R.}~\bibnamefont{Whitworth}},
  \emph{\bibinfo{title}{Physics of Ice}} (\bibinfo{publisher}{OUP Oxford},
  \bibinfo{year}{1999}), ISBN \bibinfo{isbn}{9780191581342}.

\bibitem[{\citenamefont{Volkov et~al.}(2014)\citenamefont{Volkov, Artemov, and
  Pronin}}]{Volkov2014:46004}
\bibinfo{author}{\bibfnamefont{A.~A.} \bibnamefont{Volkov}},
  \bibinfo{author}{\bibfnamefont{V.~G.} \bibnamefont{Artemov}},
  \bibnamefont{and} \bibinfo{author}{\bibfnamefont{A.~V.}
  \bibnamefont{Pronin}}, \bibinfo{journal}{EPL (Europhysics Letters)}
  \textbf{\bibinfo{volume}{106}}, \bibinfo{pages}{46004}
  (\bibinfo{year}{2014}).

\bibitem[{\citenamefont{Elton and
  Fern{\'a}ndez-Serra}(2014)}]{Elton2014:124504}
\bibinfo{author}{\bibfnamefont{D.~C.} \bibnamefont{Elton}} \bibnamefont{and}
  \bibinfo{author}{\bibfnamefont{M.-V.} \bibnamefont{Fern{\'a}ndez-Serra}},
  \bibinfo{journal}{J. Chem. Phys.} \textbf{\bibinfo{volume}{140}},
  \bibinfo{pages}{124504} (\bibinfo{year}{2014}).

\bibitem[{\citenamefont{Sega and Schr\"{o}der}(2015)}]{Sega2015:1539}
\bibinfo{author}{\bibfnamefont{M.}~\bibnamefont{Sega}} \bibnamefont{and}
  \bibinfo{author}{\bibfnamefont{C.}~\bibnamefont{Schr\"{o}der}},
  \bibinfo{journal}{The J. Phys. Chem. A} \textbf{\bibinfo{volume}{119}},
  \bibinfo{pages}{1539} (\bibinfo{year}{2015}).

\bibitem[{\citenamefont{Sciortino
  et~al.}(1990{\natexlab{a}})\citenamefont{Sciortino, Geiger, and
  Stanley}}]{Sciortino1990:3452}
\bibinfo{author}{\bibfnamefont{F.}~\bibnamefont{Sciortino}},
  \bibinfo{author}{\bibfnamefont{A.}~\bibnamefont{Geiger}}, \bibnamefont{and}
  \bibinfo{author}{\bibfnamefont{H.~E.} \bibnamefont{Stanley}},
  \bibinfo{journal}{Phys. Rev. Lett.} \textbf{\bibinfo{volume}{65}},
  \bibinfo{pages}{3452} (\bibinfo{year}{1990}{\natexlab{a}}).

\bibitem[{\citenamefont{Elton and Fern{\'a}ndez-Serra}(2016)}]{Elton2016:10193}
\bibinfo{author}{\bibfnamefont{D.~C.} \bibnamefont{Elton}} \bibnamefont{and}
  \bibinfo{author}{\bibfnamefont{M.-V.} \bibnamefont{Fern{\'a}ndez-Serra}},
  \bibinfo{journal}{Nat. Comm.} \textbf{\bibinfo{volume}{7}},
  \bibinfo{pages}{10193} (\bibinfo{year}{2016}).

\bibitem[{\citenamefont{Elton}(2016)}]{EltonThesis}
\bibinfo{author}{\bibfnamefont{D.~C.} \bibnamefont{Elton}}, Ph.D. thesis,
  \bibinfo{school}{Stony Brook University} (\bibinfo{year}{2016}).

\bibitem[{\citenamefont{Barthel et~al.}(1990)\citenamefont{Barthel, Bachhuber,
  Buchner, and Hetzenauer}}]{Barthel1990:369}
\bibinfo{author}{\bibnamefont{Barthel}},
  \bibinfo{author}{\bibfnamefont{K.}~\bibnamefont{Bachhuber}},
  \bibinfo{author}{\bibfnamefont{R.}~\bibnamefont{Buchner}}, \bibnamefont{and}
  \bibinfo{author}{\bibfnamefont{H.}~\bibnamefont{Hetzenauer}},
  \bibinfo{journal}{Chem. Phys. Lett.} \textbf{\bibinfo{volume}{165}},
  \bibinfo{pages}{369 } (\bibinfo{year}{1990}).

\bibitem[{\citenamefont{Peacock}(2009)}]{Peakcock2009:205501}
\bibinfo{author}{\bibfnamefont{J.~R.} \bibnamefont{Peacock}},
  \bibinfo{journal}{J. Phys. D: Appl. Phys.} \textbf{\bibinfo{volume}{42}},
  \bibinfo{pages}{205501} (\bibinfo{year}{2009}).

\bibitem[{\citenamefont{Sato and Buchner}(2004)}]{Sato2004:5007}
\bibinfo{author}{\bibfnamefont{T.}~\bibnamefont{Sato}} \bibnamefont{and}
  \bibinfo{author}{\bibfnamefont{R.}~\bibnamefont{Buchner}},
  \bibinfo{journal}{J. Phys. Chem. A} \textbf{\bibinfo{volume}{108}},
  \bibinfo{pages}{5007} (\bibinfo{year}{2004}).

\bibitem[{\citenamefont{Moller et~al.}(2009)\citenamefont{Moller, Cooke,
  Tanaka, and Jepsen}}]{Moller2009:A113}
\bibinfo{author}{\bibfnamefont{U.}~\bibnamefont{Moller}},
  \bibinfo{author}{\bibfnamefont{D.~G.} \bibnamefont{Cooke}},
  \bibinfo{author}{\bibfnamefont{K.}~\bibnamefont{Tanaka}}, \bibnamefont{and}
  \bibinfo{author}{\bibfnamefont{P.~U.} \bibnamefont{Jepsen}},
  \bibinfo{journal}{J. Opt. Soc. Am. B} \textbf{\bibinfo{volume}{26}},
  \bibinfo{pages}{A113} (\bibinfo{year}{2009}).

\bibitem[{\citenamefont{Yadaa et~al.}(2008)\citenamefont{Yadaa, Nagaia, and
  Tanaka}}]{Yada2008:166}
\bibinfo{author}{\bibfnamefont{H.}~\bibnamefont{Yadaa}},
  \bibinfo{author}{\bibfnamefont{M.}~\bibnamefont{Nagaia}}, \bibnamefont{and}
  \bibinfo{author}{\bibfnamefont{K.}~\bibnamefont{Tanaka}},
  \bibinfo{journal}{Chem. Phys. Lett.} \textbf{\bibinfo{volume}{464}},
  \bibinfo{pages}{166 } (\bibinfo{year}{2008}).

\bibitem[{\citenamefont{Fukasawa et~al.}(2005)\citenamefont{Fukasawa, Sato,
  Watanabe, Hama, Kunz, and Buchner}}]{Fukasawa2005:197802}
\bibinfo{author}{\bibfnamefont{T.}~\bibnamefont{Fukasawa}},
  \bibinfo{author}{\bibfnamefont{T.}~\bibnamefont{Sato}},
  \bibinfo{author}{\bibfnamefont{J.}~\bibnamefont{Watanabe}},
  \bibinfo{author}{\bibfnamefont{Y.}~\bibnamefont{Hama}},
  \bibinfo{author}{\bibfnamefont{W.}~\bibnamefont{Kunz}}, \bibnamefont{and}
  \bibinfo{author}{\bibfnamefont{R.}~\bibnamefont{Buchner}},
  \bibinfo{journal}{Phys. Rev. Lett.} \textbf{\bibinfo{volume}{95}},
  \bibinfo{pages}{197802} (\bibinfo{year}{2005}).

\bibitem[{\citenamefont{Venables and Schmuttenmaer}(1998)}]{Venables1998:4935}
\bibinfo{author}{\bibfnamefont{D.~S.} \bibnamefont{Venables}} \bibnamefont{and}
  \bibinfo{author}{\bibfnamefont{C.~A.} \bibnamefont{Schmuttenmaer}},
  \bibinfo{journal}{J. Chem. Phys.} \textbf{\bibinfo{volume}{108}},
  \bibinfo{pages}{4935} (\bibinfo{year}{1998}).

\bibitem[{\citenamefont{Liebe et~al.}(1991)\citenamefont{Liebe, Hufford, and
  Manabe}}]{Liebe1991:659}
\bibinfo{author}{\bibfnamefont{H.~J.} \bibnamefont{Liebe}},
  \bibinfo{author}{\bibfnamefont{G.~A.} \bibnamefont{Hufford}},
  \bibnamefont{and} \bibinfo{author}{\bibfnamefont{T.}~\bibnamefont{Manabe}},
  \bibinfo{journal}{International journal of Infrared and Millimeter Waves}
  \textbf{\bibinfo{volume}{12}}, \bibinfo{pages}{659} (\bibinfo{year}{1991}).

\bibitem[{\citenamefont{Beneduci}(2008)}]{Beneduci2008:55}
\bibinfo{author}{\bibfnamefont{A.}~\bibnamefont{Beneduci}},
  \bibinfo{journal}{J. Mol. Liq.} \textbf{\bibinfo{volume}{138}},
  \bibinfo{pages}{55 } (\bibinfo{year}{2008}).

\bibitem[{\citenamefont{Ellison}(2007)}]{ellison:1}
\bibinfo{author}{\bibfnamefont{W.~J.} \bibnamefont{Ellison}},
  \bibinfo{journal}{J. Phys. Chem. Ref. Data} \textbf{\bibinfo{volume}{36}},
  \bibinfo{pages}{1} (\bibinfo{year}{2007}).

\bibitem[{\citenamefont{Ohmine}(1995)}]{Ohmine1995:6767}
\bibinfo{author}{\bibfnamefont{I.}~\bibnamefont{Ohmine}}, \bibinfo{journal}{J.
  Phys. Chem.} \textbf{\bibinfo{volume}{99}}, \bibinfo{pages}{6767}
  (\bibinfo{year}{1995}).

\bibitem[{\citenamefont{Sciortino
  et~al.}(1990{\natexlab{b}})\citenamefont{Sciortino, Poole, Stanley, and
  Havlin}}]{Sciortino1990:1686}
\bibinfo{author}{\bibfnamefont{F.}~\bibnamefont{Sciortino}},
  \bibinfo{author}{\bibfnamefont{P.~H.} \bibnamefont{Poole}},
  \bibinfo{author}{\bibfnamefont{H.~E.} \bibnamefont{Stanley}},
  \bibnamefont{and} \bibinfo{author}{\bibfnamefont{S.}~\bibnamefont{Havlin}},
  \bibinfo{journal}{Phys. Rev. Lett.} \textbf{\bibinfo{volume}{64}},
  \bibinfo{pages}{1686} (\bibinfo{year}{1990}{\natexlab{b}}).

\bibitem[{\citenamefont{Heyden et~al.}(2010)\citenamefont{Heyden, Sun, Funkner,
  Mathias, Forbert, Havenith, and Marx}}]{Heyden2010:12068}
\bibinfo{author}{\bibfnamefont{M.}~\bibnamefont{Heyden}},
  \bibinfo{author}{\bibfnamefont{J.}~\bibnamefont{Sun}},
  \bibinfo{author}{\bibfnamefont{S.}~\bibnamefont{Funkner}},
  \bibinfo{author}{\bibfnamefont{G.}~\bibnamefont{Mathias}},
  \bibinfo{author}{\bibfnamefont{H.}~\bibnamefont{Forbert}},
  \bibinfo{author}{\bibfnamefont{M.}~\bibnamefont{Havenith}}, \bibnamefont{and}
  \bibinfo{author}{\bibfnamefont{D.}~\bibnamefont{Marx}},
  \bibinfo{journal}{Proc. Natl. Acad. Sci. USA} \textbf{\bibinfo{volume}{107}},
  \bibinfo{pages}{12068} (\bibinfo{year}{2010}).

\bibitem[{\citenamefont{Cho et~al.}(1994)\citenamefont{Cho, Fleming, Saito,
  Ohmine, and Stratt}}]{Cho1994:6672}
\bibinfo{author}{\bibfnamefont{M.}~\bibnamefont{Cho}},
  \bibinfo{author}{\bibfnamefont{G.~R.} \bibnamefont{Fleming}},
  \bibinfo{author}{\bibfnamefont{S.}~\bibnamefont{Saito}},
  \bibinfo{author}{\bibfnamefont{I.}~\bibnamefont{Ohmine}}, \bibnamefont{and}
  \bibinfo{author}{\bibfnamefont{R.~M.} \bibnamefont{Stratt}},
  \bibinfo{journal}{J. Chem. Phys.} \textbf{\bibinfo{volume}{100}},
  \bibinfo{pages}{6672} (\bibinfo{year}{1994}).

\bibitem[{\citenamefont{Poley}(1955)}]{Poley1955:337}
\bibinfo{author}{\bibfnamefont{J.}~\bibnamefont{Poley}}, \bibinfo{journal}{App.
  Sci. Res. B} \textbf{\bibinfo{volume}{4}}, \bibinfo{pages}{337}
  (\bibinfo{year}{1955}).

\bibitem[{\citenamefont{Hill}(1971)}]{Hill1971:2322}
\bibinfo{author}{\bibfnamefont{N.~E.} \bibnamefont{Hill}}, \bibinfo{journal}{J.
  Phys. C.: Solid State Phys.} \textbf{\bibinfo{volume}{4}},
  \bibinfo{pages}{2322} (\bibinfo{year}{1971}).

\bibitem[{\citenamefont{Nora E.~Hill}(1963)}]{H69}
\bibinfo{author}{\bibfnamefont{A.~P. M.~D.} \bibnamefont{Nora E.~Hill},
  \bibfnamefont{Worth E.~Vaughan}}, \emph{\bibinfo{title}{Dielectric Properties
  and Molecular Behaviour}} (\bibinfo{publisher}{Van Norstrand Reinhold
  Company}, \bibinfo{address}{New York}, \bibinfo{year}{1963}).

\bibitem[{\citenamefont{Bagchi and Chandra}(1993)}]{Bagchi1993:133}
\bibinfo{author}{\bibfnamefont{B.}~\bibnamefont{Bagchi}} \bibnamefont{and}
  \bibinfo{author}{\bibfnamefont{A.}~\bibnamefont{Chandra}},
  \bibinfo{journal}{Chem. Phys.} \textbf{\bibinfo{volume}{173}},
  \bibinfo{pages}{133 } (\bibinfo{year}{1993}).

\bibitem[{\citenamefont{Ferraro and Basile}(1978)}]{ferraro1978}
\bibinfo{author}{\bibfnamefont{R.}~\bibnamefont{Ferraro}} \bibnamefont{and}
  \bibinfo{author}{\bibfnamefont{L.}~\bibnamefont{Basile}},
  \emph{\bibinfo{title}{Fourier Transform Infrared Spectra: Applications to
  Chem. Systems}}, vol. \bibinfo{volume}{v. 1} (\bibinfo{publisher}{Elsevier
  Science}, \bibinfo{year}{1978}), ISBN \bibinfo{isbn}{9780323140171}.

\bibitem[{\citenamefont{Speedy and Angell}(1976)}]{speedy:851}
\bibinfo{author}{\bibfnamefont{R.~J.} \bibnamefont{Speedy}} \bibnamefont{and}
  \bibinfo{author}{\bibfnamefont{C.~A.} \bibnamefont{Angell}},
  \bibinfo{journal}{J. Chem. Phys.} \textbf{\bibinfo{volume}{65}},
  \bibinfo{pages}{851} (\bibinfo{year}{1976}).

\bibitem[{\citenamefont{Starr et~al.}(1999)\citenamefont{Starr, Nielsen, and
  Stanley}}]{PhysRevLett.82.2294}
\bibinfo{author}{\bibfnamefont{F.~W.} \bibnamefont{Starr}},
  \bibinfo{author}{\bibfnamefont{J.~K.} \bibnamefont{Nielsen}},
  \bibnamefont{and} \bibinfo{author}{\bibfnamefont{H.~E.}
  \bibnamefont{Stanley}}, \bibinfo{journal}{Phys. Rev. Lett.}
  \textbf{\bibinfo{volume}{82}}, \bibinfo{pages}{2294} (\bibinfo{year}{1999}).

\bibitem[{\citenamefont{Pirc and Blinc}(2007)}]{Pirc:020101}
\bibinfo{author}{\bibfnamefont{R.}~\bibnamefont{Pirc}} \bibnamefont{and}
  \bibinfo{author}{\bibfnamefont{R.}~\bibnamefont{Blinc}},
  \bibinfo{journal}{Phys. Rev. B} \textbf{\bibinfo{volume}{76}},
  \bibinfo{pages}{020101} (\bibinfo{year}{2007}).

\bibitem[{\citenamefont{Bokov et~al.}(1999)\citenamefont{Bokov, Leshchenko,
  Malitskaya, and Raevski}}]{Bokov:4899}
\bibinfo{author}{\bibfnamefont{A.~A.} \bibnamefont{Bokov}},
  \bibinfo{author}{\bibfnamefont{M.~A.} \bibnamefont{Leshchenko}},
  \bibinfo{author}{\bibfnamefont{M.~A.} \bibnamefont{Malitskaya}},
  \bibnamefont{and} \bibinfo{author}{\bibfnamefont{I.~P.}
  \bibnamefont{Raevski}}, \bibinfo{journal}{J. Phys.: Cond. Mat.}
  \textbf{\bibinfo{volume}{11}}, \bibinfo{pages}{4899} (\bibinfo{year}{1999}).

\bibitem[{\citenamefont{Vilgis}(1990)}]{Vilgis:3667}
\bibinfo{author}{\bibfnamefont{T.~A.} \bibnamefont{Vilgis}},
  \bibinfo{journal}{journal of Physics: Condensed Matter}
  \textbf{\bibinfo{volume}{2}}, \bibinfo{pages}{3667} (\bibinfo{year}{1990}).

\bibitem[{\citenamefont{Giovambattista
  et~al.}(2003)\citenamefont{Giovambattista, Buldyrev, Starr, and
  Stanley}}]{Giovambattista2003:085506}
\bibinfo{author}{\bibfnamefont{N.}~\bibnamefont{Giovambattista}},
  \bibinfo{author}{\bibfnamefont{S.~V.} \bibnamefont{Buldyrev}},
  \bibinfo{author}{\bibfnamefont{F.~W.} \bibnamefont{Starr}}, \bibnamefont{and}
  \bibinfo{author}{\bibfnamefont{H.~E.} \bibnamefont{Stanley}},
  \bibinfo{journal}{Phys. Rev. Lett.} \textbf{\bibinfo{volume}{90}},
  \bibinfo{pages}{085506} (\bibinfo{year}{2003}).

\bibitem[{\citenamefont{Tagantsev}(1994)}]{Tagantsev1994:1100}
\bibinfo{author}{\bibfnamefont{A.~K.} \bibnamefont{Tagantsev}},
  \bibinfo{journal}{Phys. Rev. Lett.} \textbf{\bibinfo{volume}{72}},
  \bibinfo{pages}{1100} (\bibinfo{year}{1994}).

\bibitem[{\citenamefont{Abascal and Vega}(2005)}]{abascal:234505}
\bibinfo{author}{\bibfnamefont{J.~L.~F.} \bibnamefont{Abascal}}
  \bibnamefont{and} \bibinfo{author}{\bibfnamefont{C.}~\bibnamefont{Vega}},
  \bibinfo{journal}{J. Chem. Phys.} \textbf{\bibinfo{volume}{123}},
  \bibinfo{eid}{234505} (\bibinfo{year}{2005}).

\bibitem[{\citenamefont{Sch{\"a}fer et~al.}(1996)\citenamefont{Sch{\"a}fer,
  Sternin, Stannarius, Arndt, and Kremer}}]{Schafer:2177}
\bibinfo{author}{\bibfnamefont{H.}~\bibnamefont{Sch{\"a}fer}},
  \bibinfo{author}{\bibfnamefont{E.}~\bibnamefont{Sternin}},
  \bibinfo{author}{\bibfnamefont{R.}~\bibnamefont{Stannarius}},
  \bibinfo{author}{\bibfnamefont{M.}~\bibnamefont{Arndt}}, \bibnamefont{and}
  \bibinfo{author}{\bibfnamefont{F.}~\bibnamefont{Kremer}},
  \bibinfo{journal}{Phys. Rev. Lett.} \textbf{\bibinfo{volume}{76}},
  \bibinfo{pages}{2177} (\bibinfo{year}{1996}).

\bibitem[{\citenamefont{Dias}(1996)}]{Dias:14212}
\bibinfo{author}{\bibfnamefont{C.~J.} \bibnamefont{Dias}},
  \bibinfo{journal}{Phys. Rev. B} \textbf{\bibinfo{volume}{53}},
  \bibinfo{pages}{14212} (\bibinfo{year}{1996}).

\bibitem[{\citenamefont{Weese}(1992)}]{Weese1992:99}
\bibinfo{author}{\bibfnamefont{J.}~\bibnamefont{Weese}},
  \bibinfo{journal}{Computer Physics Communications}
  \textbf{\bibinfo{volume}{69}}, \bibinfo{pages}{99 } (\bibinfo{year}{1992}).

\bibitem[{\citenamefont{Tuncer and Macdonald}(2006)}]{Tuncer2006:074106}
\bibinfo{author}{\bibfnamefont{E.}~\bibnamefont{Tuncer}} \bibnamefont{and}
  \bibinfo{author}{\bibfnamefont{J.~R.} \bibnamefont{Macdonald}},
  \bibinfo{journal}{J. Appl. Phys.} \textbf{\bibinfo{volume}{99}},
  \bibinfo{eid}{074106} (\bibinfo{year}{2006}).

\bibitem[{\citenamefont{Ktitorov}(2003)}]{Jang2003:956}
\bibinfo{author}{\bibfnamefont{S.}~\bibnamefont{Ktitorov}},
  \bibinfo{journal}{Technical Physics Letters} \textbf{\bibinfo{volume}{29}},
  \bibinfo{pages}{956} (\bibinfo{year}{2003}).

\bibitem[{\citenamefont{Sheppard and Grant}(1974)}]{Sheppard1974:61}
\bibinfo{author}{\bibfnamefont{R.}~\bibnamefont{Sheppard}} \bibnamefont{and}
  \bibinfo{author}{\bibfnamefont{E.}~\bibnamefont{Grant}},
  \bibinfo{journal}{Advances in Molecular Relaxation Processes}
  \textbf{\bibinfo{volume}{6}}, \bibinfo{pages}{61 } (\bibinfo{year}{1974}).

\bibitem[{\citenamefont{Barker}(1975)}]{Barker1975:4071}
\bibinfo{author}{\bibfnamefont{A.~S.} \bibnamefont{Barker}},
  \bibinfo{journal}{Phys. Rev. B} \textbf{\bibinfo{volume}{12}},
  \bibinfo{pages}{4071} (\bibinfo{year}{1975}).

\bibitem[{\citenamefont{Puzenko et~al.}(2005)\citenamefont{Puzenko, Hayashi,
  Ryabov, Balin, Feldman, Kaatze, and Behrends}}]{Puzenko2005:6031}
\bibinfo{author}{\bibfnamefont{A.}~\bibnamefont{Puzenko}},
  \bibinfo{author}{\bibfnamefont{Y.}~\bibnamefont{Hayashi}},
  \bibinfo{author}{\bibfnamefont{Y.~E.} \bibnamefont{Ryabov}},
  \bibinfo{author}{\bibfnamefont{I.}~\bibnamefont{Balin}},
  \bibinfo{author}{\bibfnamefont{Y.}~\bibnamefont{Feldman}},
  \bibinfo{author}{\bibfnamefont{U.}~\bibnamefont{Kaatze}}, \bibnamefont{and}
  \bibinfo{author}{\bibfnamefont{R.}~\bibnamefont{Behrends}},
  \bibinfo{journal}{J. Phys. Chem. B} \textbf{\bibinfo{volume}{109}},
  \bibinfo{pages}{6031} (\bibinfo{year}{2005}).

\bibitem[{\citenamefont{Segelstein}(1981)}]{Segelstein}
\bibinfo{author}{\bibfnamefont{D.}~\bibnamefont{Segelstein}}, Master's thesis,
  \bibinfo{school}{University of Missouri, Kansas City} (\bibinfo{year}{1981}).

\bibitem[{\citenamefont{Brendel and Bormann}(1992)}]{Brendel1992:1}
\bibinfo{author}{\bibfnamefont{R.}~\bibnamefont{Brendel}} \bibnamefont{and}
  \bibinfo{author}{\bibfnamefont{D.}~\bibnamefont{Bormann}},
  \bibinfo{journal}{J. Appl. Phys.} \textbf{\bibinfo{volume}{71}},
  \bibinfo{pages}{1} (\bibinfo{year}{1992}).

\end{thebibliography}

\end{document}